\documentclass{pasj00}

\newcommand{\bm}[1]{\mbox{\boldmath$#1$}}

\begin{document}
\SetRunningHead{T. Hamana et al.}{Anisotropic PSF of Suprime-Cam}
\Received{2013/4/18}
\Accepted{2013/6/25}
\Published{2013/10/25}

\title{Toward understanding the anisotropic point spread function of Suprime-Cam
  and its impact on cosmic shear measurement}

\author{Takashi \textsc{Hamana}\altaffilmark{1}, 
Satoshi \textsc{Miyazaki}\altaffilmark{1},
Yuki \textsc{Okura}\altaffilmark{1},
Tomohiro \textsc{Okamura}\altaffilmark{2},
Toshifumi \textsc{Futamase}\altaffilmark{2}}
\altaffiltext{1}{National Astronomical Observatory of Japan, Mitaka, 
Tokyo 181-8588, Japan}
\altaffiltext{2}{Astronomical Institute, Tohoku University, Sendai, 980-8578, Japan}


\maketitle

\begin{abstract}
We examined the anisotropic point spread function (PSF) of Suprime-Cam
data utilizing the dense star field data. 
We decompose the PSF ellipticities into three components, the optical
aberration, atmospheric turbulence and chip-misalignment in an
empirical manner, and evaluate the amplitude of each component.
We then tested a standard method for correcting the PSF ellipticities
used in weak lensing analysis against mock simulation.
We found that, for long-exposure data, the optical aberration has the
largest contribution to the PSF ellipticities, which could be modeled well
by a simple analytic function based on the lowest-order aberration theory.
The statistical properties of PSF ellipticities resulting from the
atmospheric turbulence are investigated by using the numerical
simulation. 
The simulation results are in a reasonable
agreement with the observed data.
It follows from those findings that the spatial variation of PSF
ellipticities consists of two components; one is a smooth and
parameterizable component arising from the optical PSF, and the other is
a non-smooth and stochastic component resulting from the atmospheric PSF.
The former can be well corrected by the standard correction method with 
polynomial fitting function.
However, for the later, its correction is affected by the common limitation
caused by sparse sampling of PSFs due to a limited number of stars.
We also examine effects of the residual PSF anisotropies on
Suprime-Cam cosmic shear data (5.6-degree$^2$ of $i'$-band data).
We found that the shape and amplitude of B-mode shear variance are 
broadly consistent with those of the residual PSF ellipticities measured
from the dense star field data.
This indicates that most of the sources of residual systematic are
understood, which is an important step for cosmic shear
statistics to be a practical tool of the precision cosmology.
\end{abstract}

%
%
\section{Introduction}
Weak gravitational lensing has now became a unique and practical tool to 
probe dark matter distribution irrespective of its relation to luminous
objects, thanks to great progress in technique for weak lensing
analysis as well as in a wide field imaging instrument (for reviews see
Mellier 1999; Bartelmann \& Schneider 2001; Refregier 2003; Hoekstra \&
Jain 2008; Munshi et al. 2008). 
Measuring the statistics of weak lensing shear, also known as the
cosmic shear statistics, has been recognized as one of most powerful
probes of cosmological parameters and is employed as the primary
science application by many on-going/future wide field survey projects
(e.g., the Panoramic Survey Telescope \& Rapid Response System (Pan-STARRS); 
Dark Energy Survey (DES); 
Hyper SuprimeCam survey;
the KIlo-Degree Survey (KIDS);
the Large Synoptic Survey Telescope (LSST);
Wide Field Infrared Survey Telescope (WFIRST);
Euclid mission). 
It should be, however, noted that the full potential of cosmic shear
statistics to place a constraint on the cosmological parameters is
archived only if systematic errors are reduced down to a required level
(Huterer et al. 2006; Cropper et al. 2013; Massey et al. 2013).   

In the process of weak lensing analysis, the correction for the point
spread function (PSF) is of fundamental importance, because it has two
serious effects on any galaxy shape measurement:
One is the circularization of the images, which systematically lowers the
lensing shear. The other is a coherent deformation of the images
caused by anisotropy of the PSF, which may mimic a weak lensing shear. 
Much effort has been made to develop techniques and software to
correct the PSF effects, and the weak lensing community has conducted blind
tests using mock data to evaluate their performance (Heymans et
al. 2006a; Massey et al. 2007; Bridle et al. 2010; Kitching et
al. 2012a,b; and references therein).
Most of the software is designed to correct the PSF, regardless of
its origins.
Major sources of PSF, recognized so far, except for diffraction,
include the atmospheric turbulence, optical aberration, the
misalignment of CCD chips on a focal plane and pixelization. 
An alternative and complementary approach to this issue is to investigate
the properties of the PSF for a specific instrument, while focusing
on a specific cause(s), which may help to optimize the PSF correction scheme
and to understand a level of residual systematics (Hoekstra 2004; Jarvis
\& Jain 2004; Wittman 2005; Jarvis, Schechter \& Jain 2008; Jee et
al. 2007; 2011; Rhodes et al. 2007; Jee \& Tyson 2011; Heymans et al
2012; Chang et al. 2003).

The latter approach is exactly what we consider in the present paper.
The purpose of this study is twofold:
First, we look at the properties of anisotropic
PSFs of the Subaru prime focus Camera (Suprime-Cam, Miyazaki et al. 2002). 
We pay a special attention to evaluating the amplitude and spatial
correlation of PSF anisotropies caused by different sources.  
To do this, we used a set of dense star field images from which we can
sample the PSF densely, enabling us to investigate the spatial variation
of the PSF in detail.
Second, using mock simulation data of PSFs, we examined
how well the PSF anisotropy can be corrected
using the standard PSF correction scheme.

The motivation of this study concerns the existence of a non-zero B-mode in the
cosmic shear correlation functions measured from Suprime-Cam
data. The measurement was made by Hamana et al. (2003), they analyzed a 2
degree$^2$ data and found statistically significant B-mode
correlations (see \S \ref{sec:SCamBmode} for an updated but similar
result).  
Since the B-mode shear is not produced by gravitational lensing in
the standard gravity theory at the lowest order of the perturbative
treatment\footnote{The B-mode 
shear can arise from higher order terms but their expected amplitude
is much smaller than the observed signal (Schneider, van Waerbeke \&
Mellier 2002; Hilbert et al. 2009).}, its existence indicates any 
non-lensing process(es) taking place.
The origin(s) of the B-mode was unclear; it could be simply due to 
remaining systematics in the data reduction and analysis.
Also it may arise from an intrinsic alignment of galaxy shapes (e.g.,
Croft \& Metzler 2000; Catelan,
Kamionkowski \& Blandford 2001; Crittenden et al. 2002; Jing 2002;
Hirata \& Seljak 2004; Heymans et al. 2006b).
More importantly, the B-mode may be produced by a non-standard gravity
(e.g., Yamauchi, Namikawa \& Taruya 2012; 2013). 
Thus the B-mode shear, in principle, provides us with valuable
information on the gravity theory.
It is thus important to first understand the level of residual systematic in
the data analysis (see, for a similar approach, Hoekstra 2004; 	Van
Waerbeke, Mellier \& Hoekstra 2005).

The outline of this paper is as follows; in section 2, data reduction
and PSF measurements of dense star field data are described. 
The amplitude and spatial correlation of PSF anisotropies caused by
different sources are evaluated separately in section 3.
The PSF correction scheme adopted in our cosmic shear analysis is tested
against the mock simulation data in section 4.
In section 5, we assess the impact of the residual PSF anisotropy on the cosmic
shear analysis.
Finally, a summary and discussion are given in section 6.
In appendices, we summarize the properties of PSF ellipticities caused
by third-order optical aberrations (appendix 1), and describe a numerical
study of PSF ellipticities caused by atmospheric turbulence (appendix 2).

\section{Data reduction and PSF measurement of dense star field images}
\label{sec:data}

We used $i'$-band data of a dense stellar field 
taken with the Suprime-Cam on 2002/9/30, 2003/6/30 and 2003/7/1.
The field is located at the galactic coordinate of 
$(l,b)=(38\degree,-3\degree)$.
We collected 289 shots of 60 seconds exposure and 70 shots of 30 seconds
exposure from the data archive {\sc SMOKA}\footnote{\tt
  http://smoka.nao.ac.jp/}.
The seeing FWHMs of those data range from 0.46\arcsec
to 0.88\arcsec with a median of 0.63\arcsec.

Each CCD data was reduced using the {\sc SDFred}\footnote{In the
  process of the correction for both
  the field distortion and differential atmospheric dispersion, the
  bi-cubic resampling scheme was implemented to suppress the aliasing effect
  (Hamana \& Miyazaki 2008).} software (Yagi et al. 2002; Ouchi et al. 2004).
Note that we conservatively used data only within 15 arcmin radius
from the field center of Suprime-Cam, because at the outside of that,
the point spread function (PSF) becomes elongated significantly, which
may make the correction for the PSF inaccurate.
Then, mosaicking of 10 CCDs was performed with {\sc SCAMP} (Bertin 2006)
and {\sc SWarp}\footnote{{\sc SWarp} was modified so that it can treat
  the bad pixel flag from the {\sc SDFred} software properly.} (Bertin
et al.~2002).
In addition to the individual exposure images, we also generated stacked
images using {\sc SWarp}.
Note that the images obtained during a same night were taken with the
same pointing (thus no dithering was made). Since differences in pointings
of images taken on different nights are very small (less than the
gaps between CCDs), no data from different CCDs was added.

Object detections were performed with {\sc SExtractor} (Bertin \& Arnouts 1996)
and {\it hfindpeaks} of {\sc IMCAT} software (Kaiser, Squires \&
Broadhurst 1995), and two catalogs were merged by matching positions with a tolerance
of 1~arcsec.
{}From the image of stars, we measured an anisotropy of the PSF.
We use the following ellipticity estimator (exactly speaking the
polarization, see Kaiser et al. 1995) to characterize the
anisotropy of PSFs:
\begin{equation}
\label{eq:e}
\bm{e}=\left({{I_{11} - I_{22}}\over {I_{11}+I_{22}}},
{{I_{12}}\over {I_{11}+I_{22}}} \right),
\end{equation}
\begin{equation}
\label{eq:I}
I_{ij}=\int d^2\theta ~ W_G(\theta) \theta_i \theta_j f(\bm{\theta}),
\end{equation}
where $f(\bm{\theta})$ is the surface brightness of an object and
$W_G(\theta)$ is the Gaussian window function. 
We used {\it getshapes} of {\sc IMCAT} for the actual computation of
these quantities.

{}From catalogs of detected objects (almost all of them are stars), we
generated two kinds of sub-catalogs by mimicking weak lensing analysis:
One is a ``star-role'' catalog which contains 700 (the number density
of 1 arcmin$^{-2}$) of the highest SN non-saturated stars. The other is
a ``galaxy-role'' catalog which contains 30,000 (42 arcmin$^{-2}$) of the
next highest SN non-saturated stars. The minimum flux SNs for star-role
and galaxy-role catalogs are sufficiently high; ${\rm SN}_{min}>330$ and
$>40$, respectively.

\section{Analysis of PSF ellipticities}
\label{sec:PSF}

\begin{figure}[t]
\begin{center}
\includegraphics[height=88mm,angle=-90]{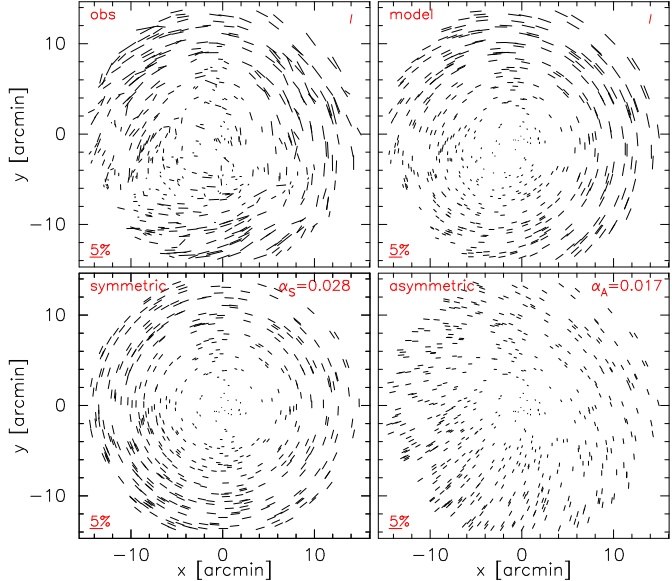}
\end{center}
\caption{Ellipticity maps, so-called {\it whisker plot}, showing 
  illustrative examples of ellipticity pattern in the Suprime-Cam data. 
A 60-second exposure image is shown.
The top-left panel shows the observed PSF ellipticites measured
from high-SN star images (``star-role'' stars), but after the mean
component being subtracted.
The mean component is displayed in the top-right corner of upper panels.
We fit the spatial variation of the observed PSFs (top-left panel) with
the optical aberration model given by eq.~(\ref{eq:PSFaberrat}).
The model consists of the symmetric (bottom-left), asymmetric
(bottom-right) and constant components. 
A sum of the first two components is shown in the top-left panel, the 
constant component is shown in the top-right corner of upper panels.
\label{fig:ESP_emap1}}
\end{figure}

\begin{figure}
\begin{center}
\includegraphics[height=88mm,angle=-90]{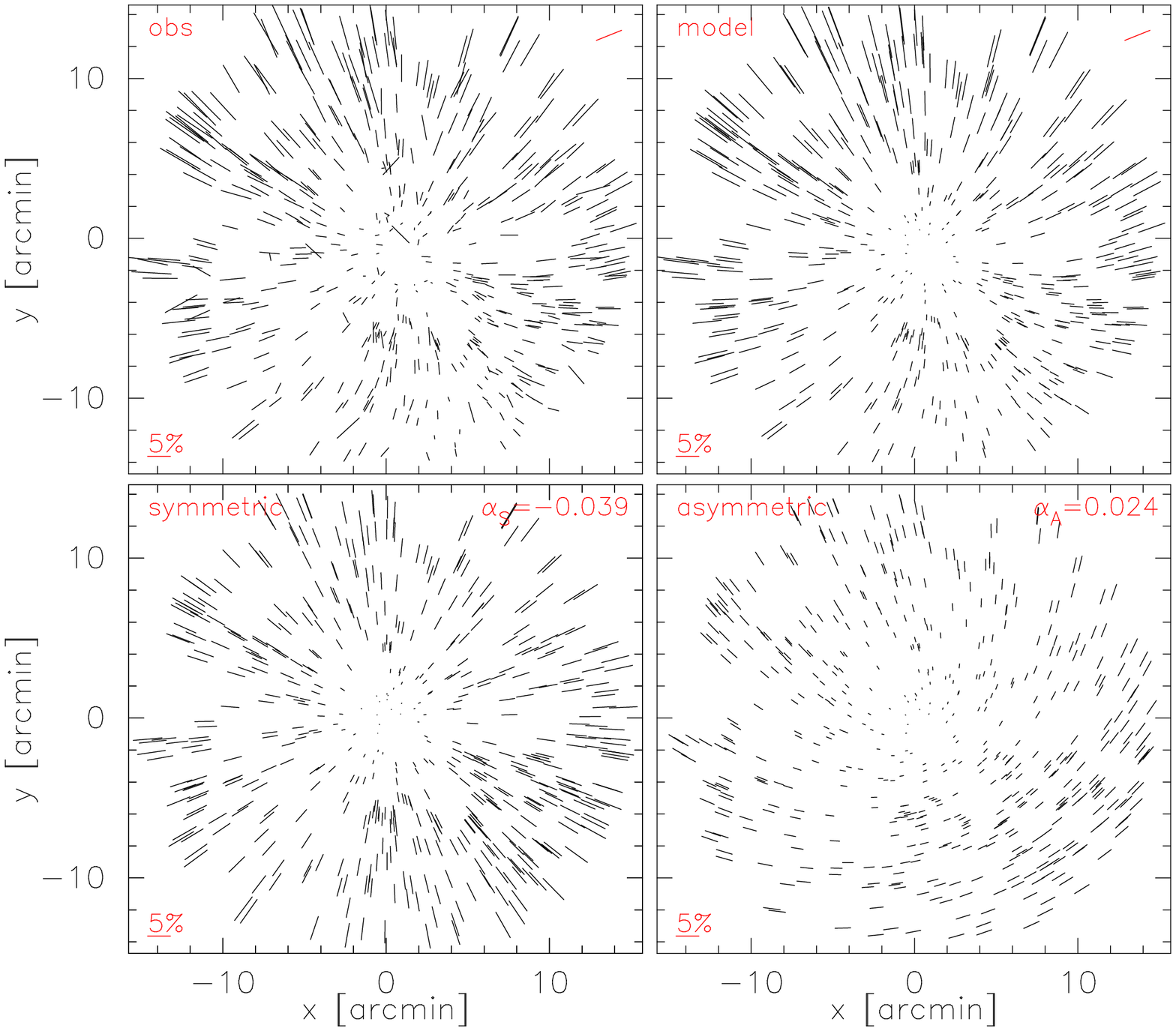}
\end{center}
\caption{Same as figure \ref{fig:ESP_emap1} but for another 60-second exposure.
\label{fig:ESP_emap2}}
\end{figure}

\begin{figure}
\begin{center}
\includegraphics[height=88mm,angle=-90]{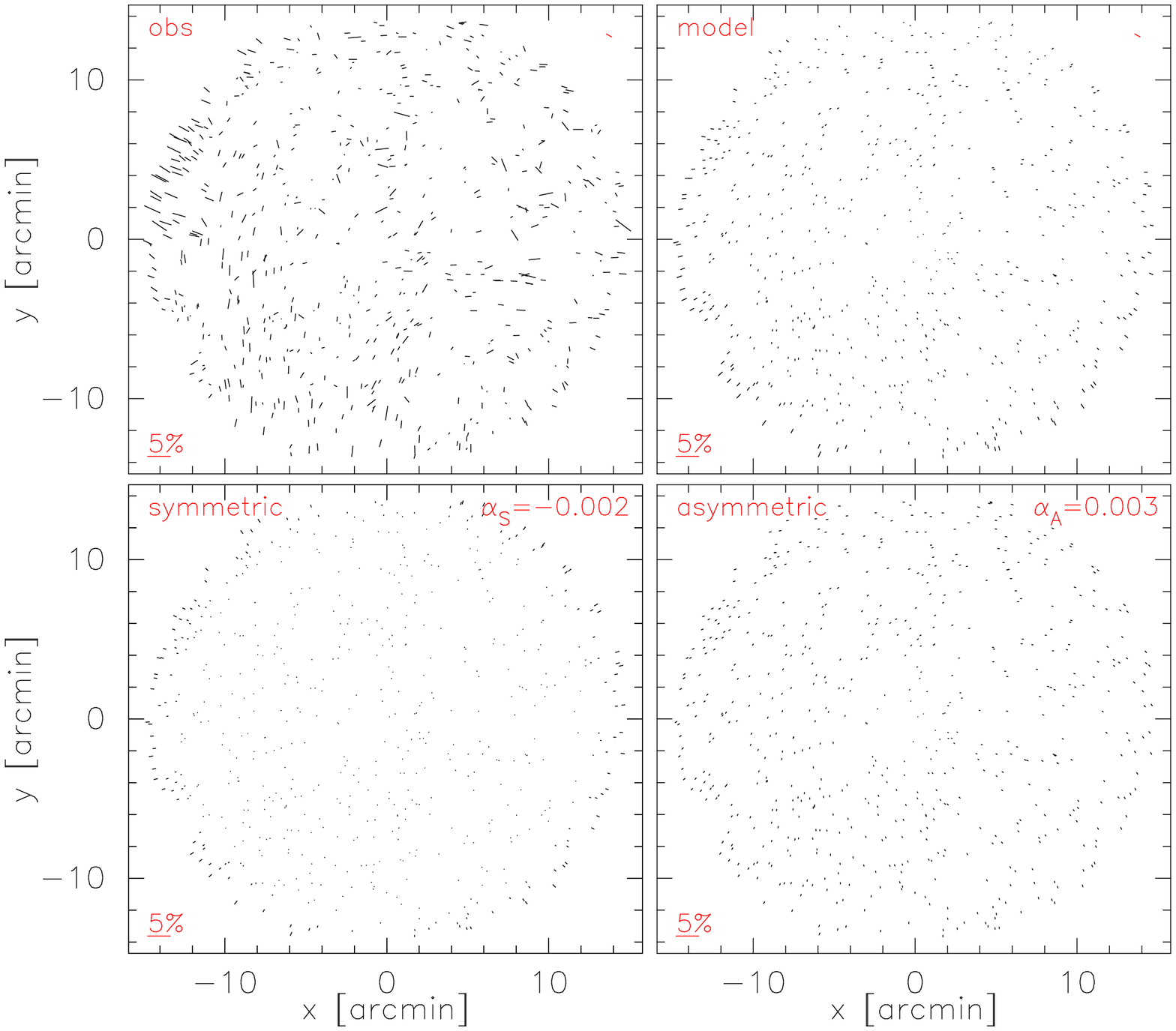}
\end{center}
\caption{Same as figure \ref{fig:ESP_emap1} but for another 60-second exposure.
\label{fig:ESP_emap3}}
\end{figure}

An anisotropy of PSF results from several sources including the
atmospheric turbulence, the optical aberration, the misalignment of CCD
chips on a focal plane, the pixelization, and tracking error of the
telescope.
Our attempt in this section is to examine the PSF on a source-by-source
basis using physically motivated models.
We evaluate the statistical properties of PSF anisotropies using a
two-point correlation function related estimator, called the E- and
B-mode aperture mass variances defined by (Schneider et al. 1998; we
follow the notation by Schneider, van Waerbeke \& Mellier 2002)
\begin{equation}
\label{eq:MapE}
\langle M_{ap}^2 \rangle(\theta) = {1 \over 2 }\int_0^{2\theta} 
{{d\phi\phi} \over {\phi^2}} 
\left[ \xi_+ (\phi) T_+\left(\phi \over \theta\right)
+\xi_- (\phi) T_-\left(\phi \over \theta\right)
\right],
\end{equation}
and
\begin{equation}
\label{eq:MapB}
\langle M_{\perp}^2 \rangle(\theta) = {1 \over 2 }\int_0^{2\theta} 
{{d\phi\phi} \over {\phi^2}} 
\left[ \xi_+ (\phi) T_+\left(\phi \over \theta\right)
-\xi_- (\phi) T_-\left(\phi \over \theta\right)
\right],
\end{equation}
where $T_+(\phi)$ and $T_-(\phi)$ are defined in Schneider et
al. (2002), and $\xi_+(\phi)$ and $\xi_-(\phi)$ are the two-point
correlation functions of the ellipticities computed by
\begin{eqnarray}
\label{eq:xi+-}
\xi_+(\theta)&=&\sum_{ij}\left(e_{i{\rm t}}e_{j{\rm
    t}}+e_{i\times}e_{j\times}\right)/N_p(\theta)\\
\xi_-(\theta)&=&\sum_{ij}\left(e_{i{\rm t}}e_{j{\rm t}}-e_{i\times}e_{j\times}\right)/N_p(\theta).
\end{eqnarray}
In the above expressions, the summation is taken over all pairs of
objects with a distance within a width of a bin considered $\theta-\Delta_{\rm bin}/2 < \phi <
\theta+\Delta_{\rm bin}/2$, $N_p(\theta)$ is the number of pairs in the
bin, and $e_t$ and $e_\times$ are the tangential and $45\degree$ rotated
ellipticity components in the frame defined by the line connecting a
pair of objects. 
Some authors adopt $\xi_{+,-}$ instead of $\langle M_{ap,\perp}^2
\rangle$ as an estimator of the spatial correlation of PSF
ellipticities and its residual, which remains after the correction (Jee
\& Tyson 2011; Chang et al 2012; 
2013). The reason for our choice of $\langle M_{ap,\perp}^2\rangle$ is
as follows. In the case of an ellipticity field with power spectrum
with a spectrum index more negative than $-2$ (this is exactly our case
as will be shown in subsequent subsections), $\xi_+$ becomes almost
flat shape, because the 
amplitude of not only the large-scale, but also the small-scale
correlation function, is dominated by the 
power on larger scales\footnote{see, e.g., the chapter 3 of the
  lecture note by N. Kaiser
  {\tt http://www.ifa.hawaii.edu/~kaiser/lectures/elements.pdf}}.
Also, even if only the larger scale PSF ellipticy correlation is
corrected, not only the large-scale but also the small-scale amplitude
decreases. 
Therefore $\xi_{+,-}$ is not suited for quantifying the spatial
correlation of our PSF ellipticity data.
On the other hand, $\langle M_{ap,\perp}^2\rangle$ is a kind of bandpass 
filtered variance, thus it is more suitable estimator for our current
purpose.

\subsection{Optical aberrations}
\label{sec:aberration}

Although PSF ellipticities arise from several causes and vary
temporally, there is a visibly identifiable component that most
likely results from optical aberrations. 
In the top-left panel of figure \ref{fig:ESP_emap1} and \ref{fig:ESP_emap2},
we show two examples of {\it whisker plots} of PSF ellipticities where
characteristic features of optical aberration can be clearly observed: 
we also show one case in figure \ref{fig:ESP_emap3} where an optical
aberration is almost invisible.  
We fit the spatial variation of PSFs with the following fitting function
of the optical PSF ellipticities, which is motivated from the
lowest-order optical aberration theory (see Appendix
\ref{appendix:aberration} for details),
\begin{eqnarray}
\label{eq:PSFaberrat}
\left(
\begin{array}{c}
  e_1\\ e_2
\end{array}
\right)
&=&
\left(
\begin{array}{c}
  c_1\\ c_2
\end{array}
\right)
+
\sum_{n_s}s_{n_s} r^{n_s} 
\left(
\begin{array}{c}
  \cos2\psi \\ \sin2\psi
\end{array}
\right)\nonumber \\
&&+
\sum_{n_a}a_{n_a} r^{n_a} 
\left(
\begin{array}{c}
  \cos(\psi-\theta_{n_a})\\ 
  \sin(\psi-\theta_{n_a}) 
\end{array}
\right),
\end{eqnarray}
where $n_s=\{1,2,3,4\}$ and $n_a=\{1,2\}$.
We determine the model parameters $s_{n_s}$, $a_{n_a}$ and $\theta_{n_a}$ by 
the standard least-square method with observed PSF ellipticities.
This model consists of the axis-symmetric, asymmetric and constant
components, which are displayed in the bottom-left panel, bottom-right
panel, and top-right corner of top-panels, respectively. 
Note that in the top-left panel, the PSF ellipticities measured from
``star-role'' stars, but after the constant component being subtracted, are
plotted.
It is clearly seen from figure \ref{fig:ESP_emap1} and \ref{fig:ESP_emap2}
that the simple three-component model can well reproduce the observed
PSFs. It may therefore be said that the optical aberration is one major
origin of the PSF anisotropy.
Note that a cause(s) of the constant component is uncertain.
It arises not only from an optical aberration (the
misalignment coma, see Appendix \ref{appendix:aberration}) but also
from, e.g., a tracking error.
Since the constant ellipticity component does not contribute to the
aperture mass variances, we do not consider it in the following
discussion.

\begin{figure}
\begin{center}
\includegraphics[width=80mm]{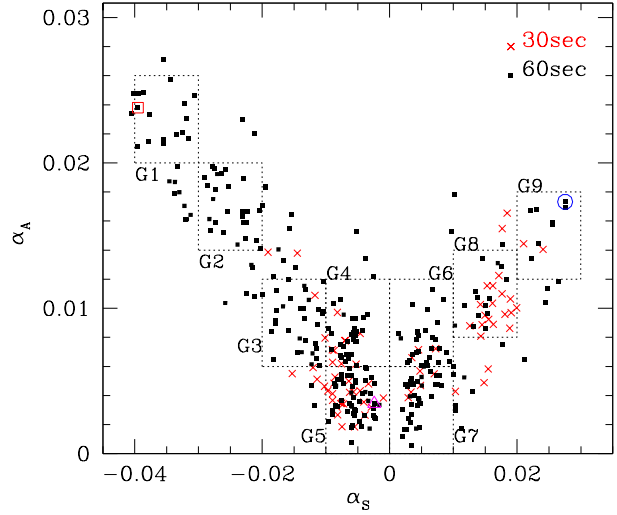}
\end{center}
\caption{Scatter plot showing the amplitude of the symmetric and
  asymmetric components of the optical aberration motivated PSF model
  for each exposure.
Crosses are for 30-second exposures, and small filled squares are for
60-second exposures.
Three examples shown in figure \ref{fig:ESP_emap1}, \ref{fig:ESP_emap2} and
\ref{fig:ESP_emap3} are marked with an open circle, open square and
open triangle, respectively.
Dotted boxes define groups used for statistical analyses.
\label{fig:gamt_prof}}
\end{figure}

Having found that the optical aberration is one major source of the PSF
ellipticities and can be well modeled with one constant and two
spatially variable components, we introduce estimators that quantify the
amplitude of the symmetric and asymmetric components:
$\alpha_S={\rm sign} \langle
e_{sym}^2 \rangle^{1/2}$ and $\alpha_A=\langle
e_{asym}^2 \rangle^{1/2}$, where the RMS scatter $\langle
e^2 \rangle^{1/2}$ is computed from the model ellipticities of 700
star-role PSFs for each exposure, and ``sign'' in $\alpha_S$ is $+$ or
$-$ for the tangentially (the mean tangential ellipticity has the plus
sign) or radially elongated cases, respectively.
The results for all exposures are plotted in
figure \ref{fig:gamt_prof}. 
It can be observed from the figure that there is a ``V-shaped'' trend
between strengths of the symmetric component and asymmetric component.
In the lowest-order aberration theory, the symmetric aberration is
caused by a shift in the focal plane position from an ideal position,
whereas the asymmetric one is caused by the decenter and/or tilt between
the axes of the focal plane and the corrector (lenses).
The result indicates that those two deviations happen in a mutually
related manner.
In figure \ref{fig:gamt_prof} one may find that there are some exposures
that are located somewhat outside of the ``V-shaped'' trend. It is
speculated from a visual inspection that the aberration model fitting of
those cases is affected due to the atmospheric PSF. In our model fitting
procedure, such contamination is unavoidable. This issue should be
noticed when one uses the model fitting for evaluating the optical PSF
component from observed PSF ellipticities.

\begin{figure}
\begin{center}
\includegraphics[width=80mm]{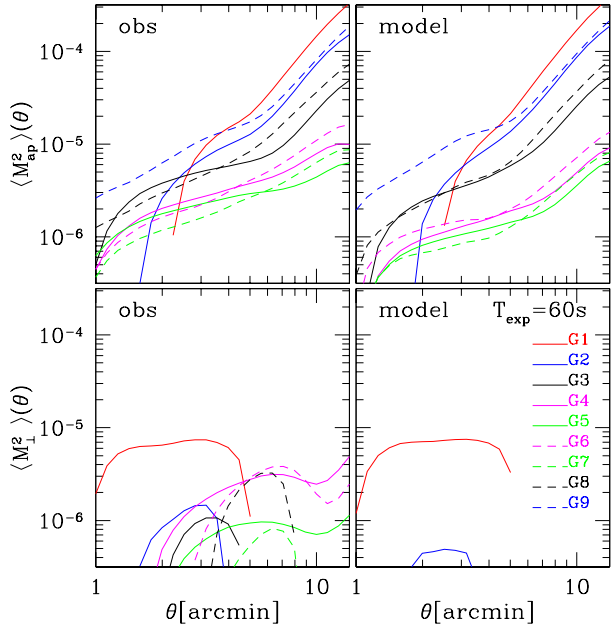}
\end{center}
\caption{Aperture mass variance of PSF ellipticities as a function
of the aperture radius. 
The top and and panels are for E- and B-mode, respectively.
The left panels show ones measured from observed data, whereas the right panels
are from the optical aberration model. Different lines are for different
groups defined in Fig \ref{fig:gamt_prof}, and the plotted aperture mass
variances are average values taken over all exposures in each group.
\label{fig:mapvar_groups}}
\end{figure}

Next we examine the statistical properties of the PSF ellipticities
using the aperture mass variance, eqs. (\ref{eq:MapE}) and
(\ref{eq:MapB}) while especially focusing on contribution from
the optical aberration. 
To do so, we classified data using $\alpha_S$ and $\alpha_A$ into 9 groups
as shown in figure \ref{fig:gamt_prof}.
We computed the average aperture mass variances for each group, and show 
the results in figure \ref{fig:mapvar_groups}.
The left panels of the figure show the results based on the observed data, from
which it is found that the E-mode variance has a larger amplitude on
larger scales, whereas in most cases B-mode variance is smaller than the
E-mode especially on larger scales.
If we assume a power-law spectrum for ellipticity power
spectrum\footnote{The aperture mass variance relates to the ellipticity
power spectrum as (Schneider et al. 1998)
$\langle M^2 \rangle (\theta) \propto \int dl~l P_e(l) I^2(l\theta)$
where $I(x)$ is the aperture function. 
Thus $\langle M^2 \rangle (\theta) \propto \theta^{-n-2}$ for a power-law power
spectrum of $P_e(l)\propto l^{n}$.}
$P_e(l)\propto l^{n}$, the measured E-mode variance indicates that 
$n\sim-3$.

The right panels of figure \ref{fig:mapvar_groups} show the aperture mass
variances computed from PSF ellipticities of the optical aberration
model.
Comparing those with observed variance, plotted in the left panels,  we
can estimate the amount of the contribution from the optical aberration to
the PSF ellipticity.
We first look at the E-mode.
It follows from the comparison between the aperture mass variance
measured from the observed ellipticities and  the model (the left and
right panels figure \ref{fig:mapvar_groups}) that for large aberration
cases such like G1-3, G8 and G9, the model variance accounts for most of
the observed one, indicating that the optical aberration is the
dominant component of the PSF ellipticities, as expected from the visual
impression (Figs.~\ref{fig:ESP_emap1} and \ref{fig:ESP_emap2}).
For small aberration cases (G4-G7), one may see the aperture mass
variances of the model are slightly smaller than the observed ones.
In the following subsections, we argue that the residual variances mostly
arise from atmospheric turbulence and chip misalignment.
Let's turn to the B-mode.
It follows from a comparison between the left and right panels that the
variances measured from the model 
are smaller than the observed ones except for G1.
This also suggests the existence of other components than the optical
component.

\subsection{CCD chip-misalignment}
\label{sec:chip}

\begin{figure}
\begin{center}
\includegraphics[height=88mm,angle=-90]{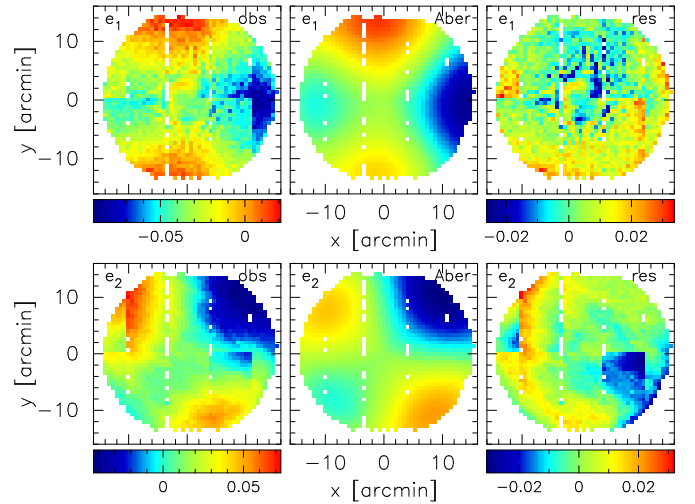}
\end{center}
\caption{Ellipticity magnitude for each component is shown: upper and
  lower panels are for $e_1$ and $e_2$, respectively.
This is taken form the same data as figure \ref{fig:ESP_emap1}.
The left panels show the observed ellipticity, whereas the central
panels show the aberration model. The color scale of the central panels
are same as that of the left panels.
The right panels show the residual ellipticity ($e_{\rm obs}-e_{\rm {opt-model}}$).
Note that the color scale of the right panels is shown in just bottom
of each plot, and are different from the other panels. 
\label{fig:emap1}}
\end{figure}

\begin{figure}[t]
\begin{center}
\includegraphics[height=88mm,angle=-90]{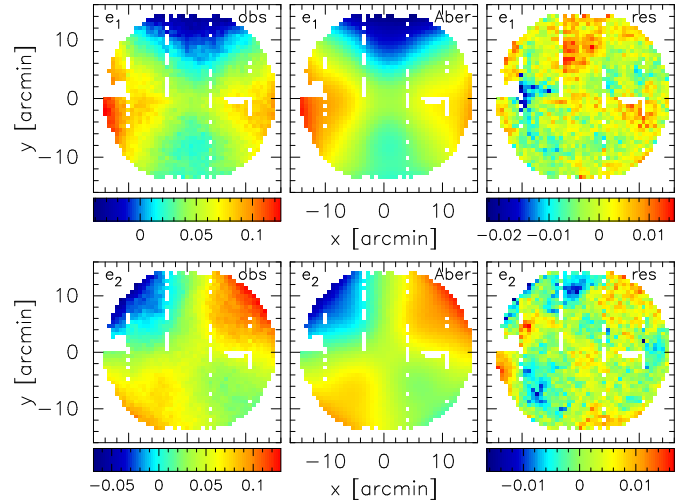}
\end{center}
\caption{Same as figure \ref{fig:emap1} but for another data shown in
  figure \ref{fig:ESP_emap2}.
\label{fig:emap2}}
\end{figure}

\begin{figure*}[t]
\begin{center}
\includegraphics[height=120mm,angle=-90]{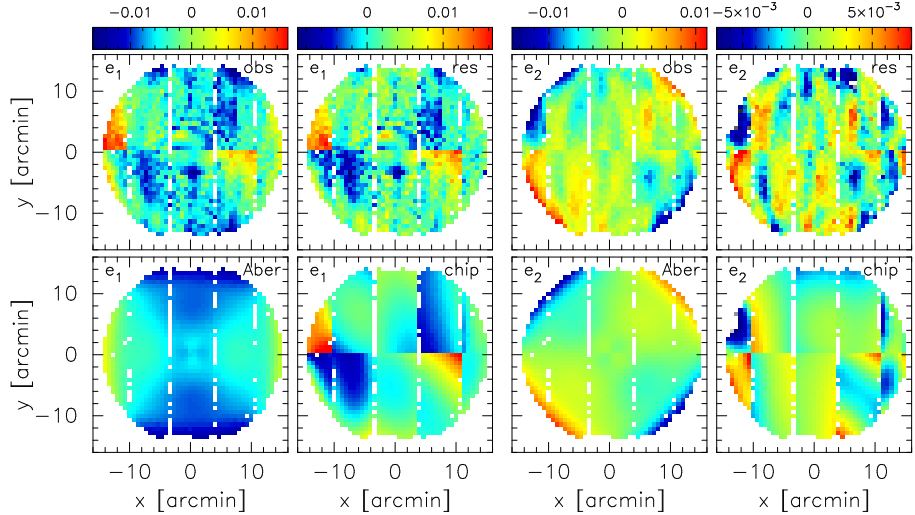}
\end{center}
\caption{Ellipticity magnitude for each component: the left and
  right set of panels are for $e_1$ and $e_2$, respectively.
Results from a coadded data of 52 60-second exposures from G5
group are shown. 
For each set of panels, top-left panel shows an observed ellipticity,
bottom-left panel is for the aberration model, top-right panel shows
the residual of those two ($e_{\rm obs}-e_{\rm{opt-model}}$), and bottom-right panel
shows the chip-misalignment model described in the subsection \ref{sec:chip}.
Note that the color scale of each column is shown in top of each column.
\label{fig:emap}}
\end{figure*}

Next, we examine the PSF ellipticities remaining after subtracting the
optical aberration component.
Let us start with a visual impression.
In figures \ref{fig:emap1} and \ref{fig:emap2}, the PSF ellipticities of
the observed data (left), the optical aberration model (center), and the 
residual of those two ($e_{\rm obs}-e_{\rm{opt-model}}$) are plotted.
Note that the mosaicked data consists of 2-row$\times$5-column CCD chips
with a masked outer region (beyond 15 arcmin radius from the field center).
In those residual plots, there are two apparent features that should be
noticed: one is discontinuities between chips, and the other is
ripple-like patterns.
The latter is a characteristic feature of PSF ellipticites caused by
atmospheric turbulence which was also observed in CFHT data (Heymans et
al. 2012), and was also found in numerical simulations (Jee \& Tyson
20011; Chang et al. 2012; 2013; also see Appendix \ref{appendix:atmos}),
which  we examine in detail in the next subsection
\ref{sec:atmospheric}. 

In this subsection, we focus on PSF ellipticities caused by misalignment
of CCD chips. To do so, we first need to create data that are minimally
affected by the optical aberration and atmospheric turbulence. 
The former could be minimized by using data with small values of
both $|\alpha_S|$ and $\alpha_A$. The latter could be suppressed by
stacking many images as the amplitude of the atmospheric PSF decreases,
with the exposure time being $\langle e^2 \rangle^{1/2}\propto T^{-1/2}$ (de
Vries et al 2007; Chang et al. 2012; Heymans et al. 2012; see also
Appendix \ref{appendix:atmos}).
Thus we generate a coadded image of 52 shots of 60-second exposures taken
from the G5 group where the effect of the optical aberration is at a minimum
level and the largest number of shots (52 shots) is contained.
The stacking was done using {\sc Swarp}, and the object detection and
ellipticity measurement were performed in the same manner as that described in
section \ref{sec:data}.

\begin{figure}
\begin{center}
\includegraphics[width=80mm]{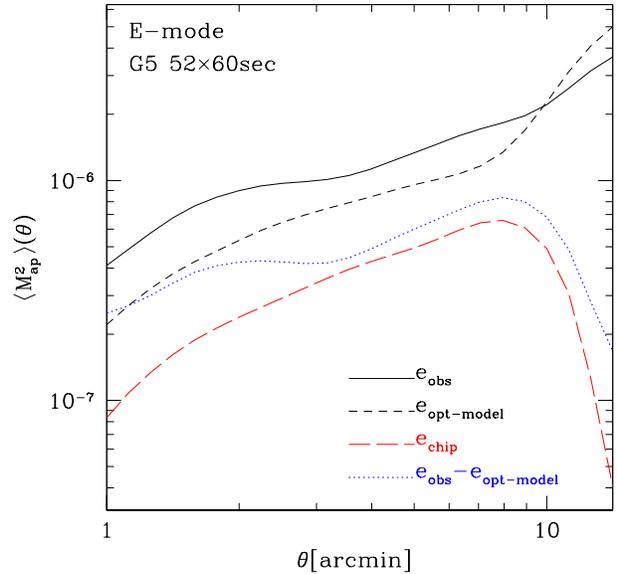}
\end{center}
\caption{E-mode aperture mass variance of PSF ellipticities measured from the coadded
  data shown in figure \ref{fig:emap}.
The solid line is for the observed PSFs, dashed line is for the aberration model,
the dotted line is for the residual of those two ($e_{\rm obs}-e_{\rm{opt-model}}$)
and the long-dashed for chip-misalignment PSF model.
\label{fig:mapvar_chip}}
\end{figure}

The PSF ellipticities of the coadded data are shown in figure \ref{fig:emap}.
One can see PSF discontinuities between the chips even in the observed PSF
(top-left panel) as well as in the residual PSFs (top-right panel). 
We attempt to extract the chip-misalignment component from the residual PSF
ellipticities by fitting the residual PSFs with the 2nd order
bi-polynomial function on the chip-by-chip basis. 
The derived PSF models are shown in the bottom-right panel of
figure \ref{fig:emap} in which it can be seen that characteristic
features of chip-misalignment PSF ellipticities (such as
discontinuities between the chips and the chip-scale smooth variation) are well
reproduced by the model. 
We measured the aperture mass variances of the PSF ellipticities (show
in figure \ref{fig:mapvar_chip}). Note that the B-mode variances are
too small to properly analyze,  and thus we do not present them.
Although our model may not extract only the chip-misalignment PSFs, but
may be affected by other sources of PSF ellipticity, it may still be
reasonable to consider that the model allows us to assess an
approximate amplitude of the chip-misalignment PSF ellipticies. 
We read from the figure \ref{fig:mapvar_chip} that the amplitude of the
chip-misalignment PSF ellipticies is roughly $\langle M_{ap}^2\rangle
\sim 10^{-7}\times\theta$.
It should, however, be noted that the chip-misalignment PSF is a
phenomenon related to the optical aberration, and thus its amplitude may
vary with deviation in the focal plane position.
Therefore the above value should be considered as a characteristic
size. 

It is also found from figure \ref{fig:mapvar_chip} that the aperture mass
variance of the residual PSFs ($e_{\rm obs}-e_{\rm {opt-model}}$) is slightly larger
than that of the chip-misalignment PSF model.
The amplitude of the excess is roughly $\langle M_{ap}^2\rangle \sim
10^{-7}$.
The origin of this excess is not clear, though it could be due to some imperfection in
the models, and/or due to atmospheric PSFs, and/or due to other causes such
as pixelization effects caused in a process of resampling (Hamana \&
Miyazaki 2008). 
If the latter is the case, it can be settled by recently developed shape 
measurements schemes that do not involve a resampling process (e.g.,
Miller et al. 2007; 2013; and Miyatake et al 2013).
Thus we leave it for future work.

\subsection{Atmospheric turbulence}
\label{sec:atmospheric}

\begin{figure}
\begin{center}
\includegraphics[width=88mm]{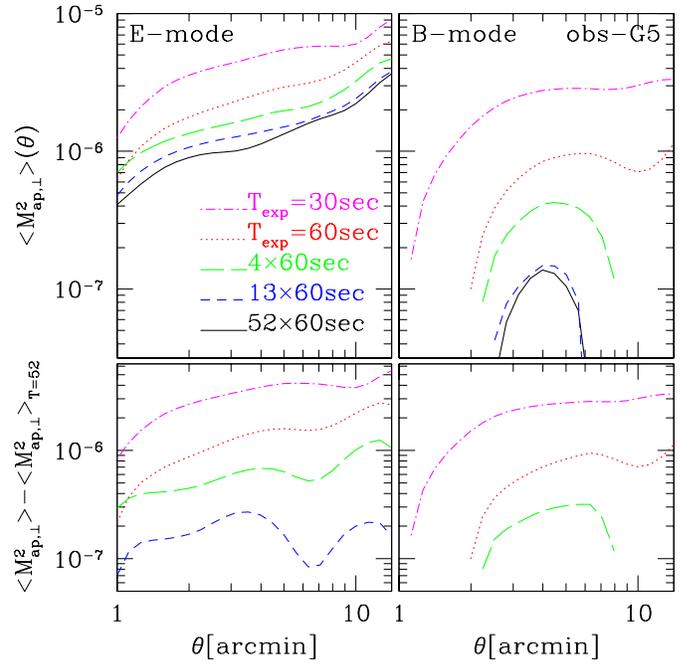}
\end{center}
\caption{{\it Top-panels}: the aperture mass variances of PSF
  ellipticities measured from
data in G5 group for different exposure times; from top to bottom,
single 30-second, 60-second exposures, coadded 4, 13 and 52 60-second exposures.
{\it Bottom-panels}: differences between aperture mass variances and
one measured from the deepest coadded data.
\label{fig:mapvar_diff}}
\end{figure}

\begin{figure}
\begin{center}
\includegraphics[width=80mm]{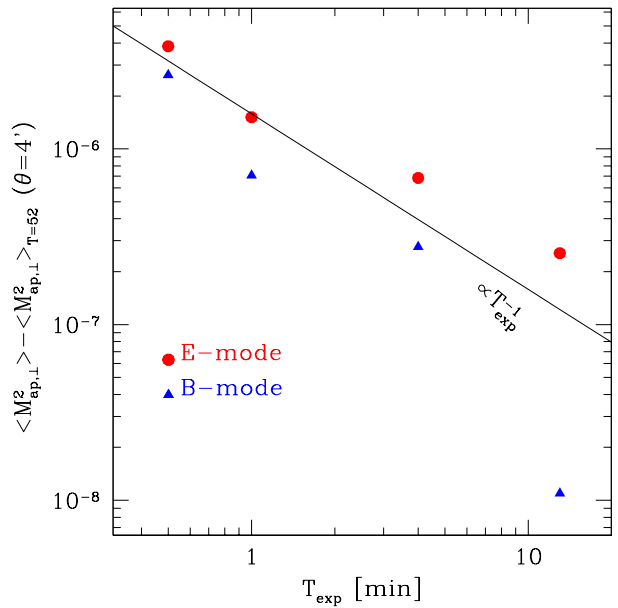}
\end{center}
\caption{Difference in the aperture mass variances from the
  52-minute exposure data at $\theta=4$ arcmin as a function of the
  exposure time.
Circles and triangles are for E- and B-mode, respectively.
The expected exposure time dependence of $\propto T_{exp}^{-1}$ (plotted
by the solid line) is clearly seen. 
\label{fig:mapvar_texp}}
\end{figure}

Let us turn to the PSF ellipticity arising from atmospheric
turbulence. The properties of the atmospheric PSF ellipticity has been
examined experimentally (Wittman 2005; Heymans et al. 2012) or using
numerical simulations (de Vries et al 2007; Jee \& Tyson 20011; Chang et
al. 2012; 2013). 
A major characteristic of the atmospheric PSF ellipticity found from
previous studies is that the amplitude of PSF ellipticities decreases
with the exposure time as $\langle e^2 \rangle^{1/2}\propto T^{-1/2}$ (de
Vries et al 2007; Chang et al. 2012; Heymans et al. 2012). 
Also, the shape of the power spectrum and aperture mass variance of the
atmospheric PSF ellipticities was investigated numerically (see appendix
\ref{appendix:atmos}).

We investigate the atmospheric PSFs using data from the G5 group.
In top-panels of figure \ref{fig:mapvar_diff}, we show the aperture mass variances
for different exposure times; from top to
bottom, single 30-second, 60-second exposures, coadded 4$\times$60, 13$\times$60
and 52$\times$60-second exposures.
As is shown in the plots, the amplitude of the aperture mass variance
decreases as the exposure time increases.
However the minimum amplitude of the E-mode is set by the optical aberration
PSFs and chip-misalignment PSFs, as shown in the last two subsections,
whereas the B-mode variance decreases down to $\langle M_{\perp}^2\rangle
\sim 10^{-7}$. This indicates that the optical and chip-misalignment PSF
largely contribute to the E-mode PSF ellipticity. 

In order to evaluate the amplitude of the atmospheric PSF ellipticities
separately from other components,
we make a working hypothesis that the amplitude of the deepest data
(52$\times 60$-sec coadded data) represents the amplitude of
(quasi-)static PSF components (such like the optical and
chip-misalignment PSFs); the difference from that gives an estimate
of atmospheric PSFs ellipticities.
In the bottom-panels of figure \ref{fig:mapvar_diff} and
figure \ref{fig:mapvar_texp}, we plot the differences, from which one finds
that the aperture mass variances have both the exposure time dependence  
and a shape similar to those obtained from
the numerical simulation (appendix \ref{appendix:atmos}).
Also, the E- and B-mode variances have similar amplitudes, as expected
from the numerical simulation.
Although the above results were derived relaying on the working
hypothesis, those nice agreements would support a reasonableness of the 
hypothesis. 
We may therefore say that we successfully captured characteristic features
of the atmospheric PSF ellipticities in the Suprime-Cam data.

\section{Testing PSF correction scheme using numerical simulations}
\label{sec:PSFcorrection}

Here we consider how well the PSF ellipticities can be corrected by
the standard correction scheme, which we describe below.
The aim of this analysis is twofold: One is to understand what sets a lower
limit of the anisotropic PSF correction; the other is to evaluate the 
amplitude of residual PSF ellipticities.
To do this, we use both the dense star field data and mock data that we
describe below.

Before going into details, we briefly describe a basic procedure concerning the
anisotropic PSF correction implemented in actual weak lensing studies so
far (see for details, Heymans et al 2006; Massey et al 2007 and references therein): 
Firstly, PSF ellipticities (or other estimators of PSF) are
measured from high-SN star images. 
In usual weak lensing analyses, a typical number density of stars used
for PSF sampling is $\sim 1$ arcmin$^{-2}$.
Secondly, the spatial variation of PSF ellipticities is modeled by
fitting the PSF ellipticities with an analytic function, such as
a polynomial. 
Then finally, an artificial shape deformation in galaxy images caused by 
the PSF is corrected using the PSF model delivered
by the analytic function.
Although an actual implementation of the above procedure depends on
technique of weak lensing analysis (Heymans et al 2006; Massey et al
2007), there is one common limitation which comes from a sparse
sampling of stars, namely, the spatial variation of PSFs on scales
smaller than the mean star separation is poorly sampled, and as a result
small-scale components in the PSF anisotropy are hardly corrected.

Our PSF correction scheme is based on the KSB algorithm (Kaiser et al
1995; Hamana et al. 2003; Heymans et al 2006). We refer the
reader to the above references for details. Here, we only describe some
details concerning implementation, which are specific to this study.
We use bi-variable polynomial function for modeling the spatial
variation of PSF ellipticites. We divide data into $2\times2$ regions
or into each CCD chip, and a PSF model is generated for each sub-region.
The order of the polynomial is set to 4th for a $2\times2$-division and
2nd or 3rd  for the chip-basis case.
The $2\times2$-division is taken because in our actual cosmic shear
analysis the PSF correction is made on a sub-region basis with a
similar area, whereas the chip-basis analysis is used to test
its ability of improving the anisotropic PSF correction. 

\begin{figure}
\begin{center}
\includegraphics[width=88mm]{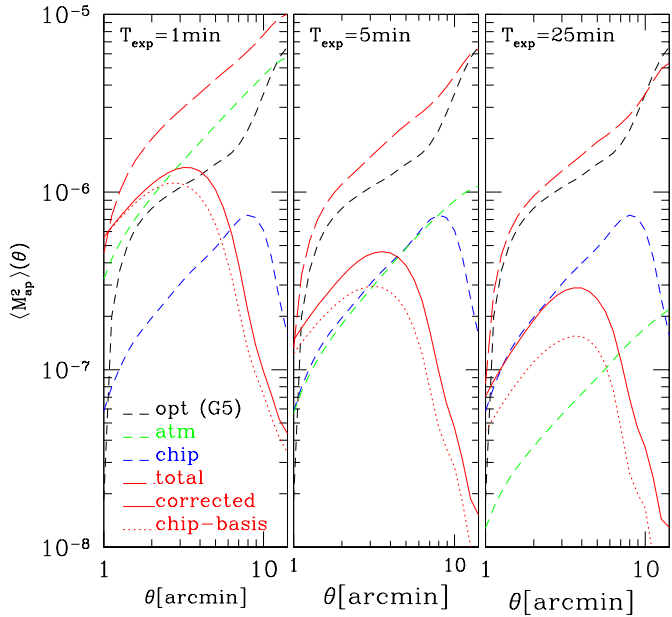}
\end{center}
\caption{E-mode aperture mass variances of PSF ellipticities before and
  after anisotropic PSF 
  correction obtained from the mock simulations. 
Dashed lines are for three input PSF models including the optical
(black), atmospheric (green) and the CCD-misalignment components.
The long-dashed lines is for total of these three components.
The solid line shows one after the correction based on
the $2\times2$-division, whereas the dotted line is for the chip-basis.
From left to right, the exposure time is 1, 5 and 25 min,
respectively. 
\label{fig:mapvar_e3_texp_chip}}
\end{figure}

Let us start with a simpler situation using mock PSF ellipticity
catalogs. The mock data have the same geometry and number of
objects as those of the dense star field data, that is 700 ``star-role''
objects and 30,000 ``galaxy-role'' objects (see section \ref{sec:data}).
Instead of generating mock image, we simply generate object catalogs. 
Each object has various items; the positions, and 3-component PSF ellipticities,
including optical, atmospheric, and CCD-misalignment components denoted
by $e_{\rm opt}$, $e_{\rm atm}$ and $e_{\rm chip}$, respectively.
The optical component is made by using the optical aberration-motivated
model, equation (\ref{eq:PSFaberrat}), introduced in Appendix
\ref{appendix:aberration}; we adopted
model parameters for G5 group in subsection \ref{sec:aberration}.
For the CCD-misalignment component, we adopt the 2nd-order polynomial
models obtained in subsection \ref{sec:chip}.
The atmospheric component is modeled by the random field with the
power spectrum having a power-law index of $-3$ based on our
finding (see section \ref{appendix:atmos} and Appendix 2).
Its amplitude is set by the empirical relation of
equations (\ref{eq:ps-scaling}) for exposure times of 1, 5, and 25 minutes.
We compute the total ellipticity by simply adding the three components,
instead of treating in the convolution manner.
This is a reasonable approximation, because each ellipticity component
is small.
Then, the anisotropic PSF correction is made for mock catalogs in the
procedure described above.

The results from a mock simulation are presented in
figure  \ref{fig:mapvar_e3_texp_chip}, from which the following three
points can be seen.
First, as mentioned above, the mean separation of stars ($\sim 1$
arcmin), which are used to model the spatial variation of PSFs, sets a
fundamental limitation on the PSF correction.
In the case of the aperture mass, its window function is most sensitive
to fluctuations on scales of $\sim 1/5$ of the aperture size (Schneider et
al. 1998),
This defines a characteristic aperture scale of $\theta\sim
5$ arcmin; on scales larger than that, the correction works very well
but on smaller scale it does very poorly (this was argued by Hoekstra et
al. 2004). 
This is especially the case for the atmospheric PSF.
Second, in spite of the above limitation, the optical components are well
corrected, even on smaller scales. The reason of this is the
smoothness of the optical component; 
as shown in subsection \ref{sec:aberration}, the spatial variation of
the optical PSFs can be well fitted by a polynomial-based model.
Thus even with a sparse sampling, it can be well-modeled with the
polynomial function. 
Third, for $2\times2$-division case, the CCD-misalignment component
combined with the optical component sets a limit on the anisotropic PSF
correction. This is a natural consequence that even if each individual
component is smooth and is modeled by a polynomial function, the mixed
PSFs may no longer be simple and thus may not be corrected by the
polynomial model. 
This can be partly avoided by adopting the chip-basis correction, as can be
seen in the right panel of figure \ref{fig:mapvar_e3_texp_chip}, though
the residual variance does not reach to the level of the atmospheric
PSF. A further improvement of the PSF correction may be achieved by
recently proposed correction schemes (e.g., Chang et al. 2012; Berg\'e et
al. 2012; Gentile, Courbin \& Meylan 2013); we leave it for future
study.  
The above three findings provide us with a clue to interpret a result
obtained from an analysis of real data.

\begin{figure}
\begin{center}
\includegraphics[width=88mm]{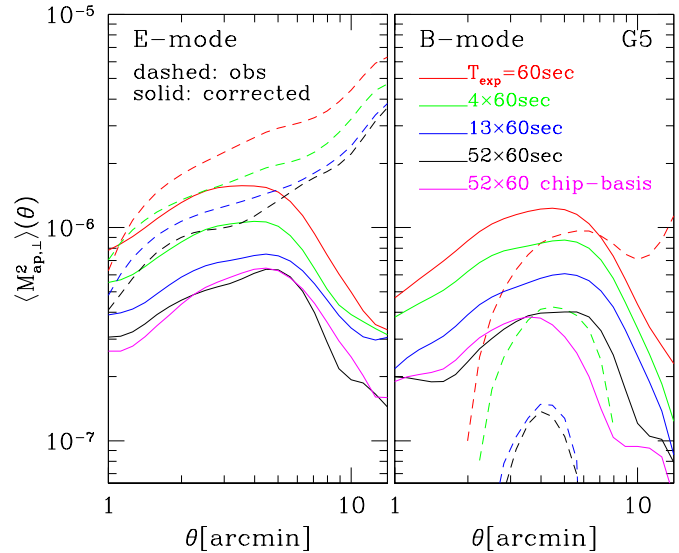}
\end{center}
\caption{E- (left) and B-mode (right) aperture mass variances of PSF
  ellipticities before
  (dashed curves) and after (solid curves) the anisotropic PSF correction. 
The results obtained from the G5 group are presented for various
exposure time. 
The PSF correction is made on a basis of $2\times2$ sub-regions except
for the magenta curve which is for the chip-basis correction.
\label{fig:mapvar_abbgroup_eb.g5cor}}
\end{figure}

Let us turn to the analysis of dense star field data.
In figure \ref{fig:mapvar_abbgroup_eb.g5cor}, the aperture mass variances
measured from the G5 group data (defined in figure  \ref{fig:gamt_prof})
are presented for different exposure times,  
where an anisotropic PSF correction was made on the basis of $2\times2$
sub-regions, except for the magenta curve, to which the chip-basis
correction was applied.
From the figure, the following four findings are obtained:
Firstly, it is found that after the PSF correction, the E- and the B-mode are
almost equally partitioned, regardless of properties of raw PSF
ellipticities. A plausible reason for this is that the E- and B-mode are
mixed up in the process of the PSF correction.
Secondly, the result of 60-sec exposure is in a good agreement with that
of the mock simulation. But at a closer look at larger scales, it is
found that the residual aperture mass variance falls less quickly than 
the mock result, indicating that the spatial variation of real PSF
ellipticities is more complex than that assumed in the mock simulation. 
Thirdly, as expected from the result of the mock simulation, the residual
aperture mass variance decreases with increasing the exposure time. 
However, the decrement is less than what can be expected from the mock
simulation. 
Finally, it is found from the comparison of two different correction
schemes for 52-coadded data that the chip-basis correction
makes an only a small improvement over the other case.
It is speculated from the last two points combined with the findings
from the mock simulation (that is, an existence of a non-smooth
component sets a lower limit of the correction on scales smaller than
mean separation of stars) that
there exists unidentified PSF ellipticity component(s) whose amplitude
is greater than the 
chip component mentioned in subsection \ref{sec:chip} (see also
figure \ref{fig:mapvar_chip}).
Although the origin of the unknown PSF component is not clear, the
above result provides us with an empirical estimate of {\it the
  ``best performance''} of the PSF correction scheme that we adopted in
this paper.

\section{Residual PSF anisotropies in Suprime-Cam cosmic shear data}
\label{sec:SCamBmode}

Here, we examine the effects of residual PSF anisotropies on
Suprime-Cam cosmic shear data. 
To do so, we measure the E- and B-mode shear aperture mass variance from
5.6-degree$^2$ Suprime-Cam deep imaging data, and compare the B-mode
shear variance with the residual PSF ellipticities measured from the
dense star field discussed in section \ref{sec:PSF}.

\subsection{Suprime-Cam data and weak lensing analysis}
\label{sec:SCamCSData}

We use the same data as those used in Hamana et al.~(2012) in which basics
of data and data analyses are described. Therefore here we only describe points
specific to this work. 
Suprime-Cam $i'$-band data were corrected from the data archive, SMOKA,
under the following three conditions: data are contiguous with at least four
pointings, the total exposure time for each pointing is longer than 1800~sec,
and the seeing full width at half-maximum (FWHM) is better than 0.65
arcsec. Four data sets (named by SXDS, COSMOS, Lockman-hole and
ELAIS-N1) meet these requirements. 
The exposure times of individual exposures range from
120 and 420 seconds.
Data reduction, mosaic stacking, and object detection were done using the
same procedure described in section \ref{sec:data}. 
The effective area after masking regions affected by bright stars is
5.57 degree$^2$. 
The depth of the coadded images varies among four fields, and it is found
that the number counts of faint galaxies are saturated at AB magnitude
of $i'=25.2 - 25.5$.  

For weak lensing measurements, we adopt the so-called KSB method
described in Kaiser et al. (2005), Luppino \& Kaiser, and Hoekstra et
al.~(1998) with some modifications being made following recent developments
(Heymans et al. 2006a), which we describe below.
Stars are selected in a standard way by
identifying the appropriate branch in the magnitude half-light radius
($r_h$) plane, along with the detection significance cut $S/N>10$. 
The number density of stars is found to be $\sim 1$~arcmin$^{-2}$ for
the four fields. 
We only use galaxies that meet the following three conditions: (i) a
detection significance of $S/N>3$ and 
$nu>10$, where $nu$ is an estimate of the peak significance given by
{\it hfindpeaks}, (ii) $r_h$ is larger than the stellar branch, and
(iii) the AB magnitude is in the range of $22 <i'<25$, where galaxy
number counts of four fields are almost the same.
The number density of resulting galaxy catalog is $\sim 23.5$
arcmin$^{-2}$ and is quite uniform among four fields.
We measure the shape parameters of objects by {\it getshapes} of {\sc
  IMCAT}. The PSF correction was done on a sub-field basis, where the
coadded data that were divided into sub-fields of about $15 \times 15$
arcmin$^2$ (approximately one fourth of the Suprime-Cam's
field-of-view). Anisotropic PSF correction was done following
the implementations by MH, CH and TS of Heymans et al.~(2006a).
The spatial variation of PSF ellipticites measured from
stars were modeled with the 4th order bi-polynomial function; also, 
the PSF model combined with the smear polarizability tensor ($P_{sm}$) of
each galaxy was used to correct the PSF ellipticities of the galaxies.
In KSB formalism, the shear (we denote by $\gamma$) is related to the
observed ellipticity through the shear polarizability tensor,
$P_\gamma$, which is evaluated by a smoothing and weighting method
developed by Van Waerbeke et al.~(2000; and see section A5 of Heymans et
al. 2006a).    

\begin{figure}
\begin{center}
\includegraphics[width=80mm]{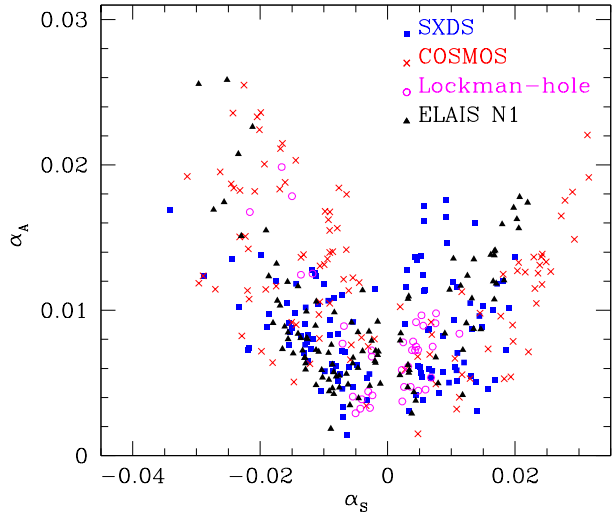}
\end{center}
\caption{Same as figure \ref{fig:gamt_prof} but for data used for the
  cosmic shear analysis.
The optical model fitting is applied to individual exposures.
The exposure time of each exposure is $120-420$ seconds.
Different symbols are for different fields.
\label{fig:eastig_emisali.scana}}
\end{figure}

Before presenting the shear aperture mass variance measured from coadded
data, it is worth looking into PSF ellipticities of individual exposures.
To do so, we generated mosaicked data of individual exposures, and selected
stars in the same procedures as described above.
The spatial variation of PSF ellipticities measured from the stars were
fitted to the optical model, equation(\ref{eq:PSFaberrat}), and the
amplitude of the symmetric and 
asymmetric components of the model were evaluated (see subsection
\ref{sec:aberration} and appendix \ref{appendix:aberration} for details).
The results are plotted in figure \ref{fig:eastig_emisali.scana}, in which 
one may see ``V-shaped'' trend similar to the results obtained from
the dense star field data (figure \ref{fig:gamt_prof}), though the scatter is
somewhat larger for this case.
This similarity implies that knowledge of the PSF anisotropy learned
from the dense star field analysis (section \ref{sec:PSF}) is applicable to
cosmic shear analyses.
It should, however, be noticed that there is one component that is
not taken into consideration in the dense star field analysis, namely 
the stacking of multiple dithered exposures.
It is obvious that the stacking of dithered exposures makes the spatial
variation of PSFs complicated, which may result in a poorer anisotropic
PSF correction. We return to this point later.

\subsection{Cosmic shear aperture mass variance}
\label{sec:cosmicshear}

\begin{figure}
\begin{center}
\includegraphics[width=88mm]{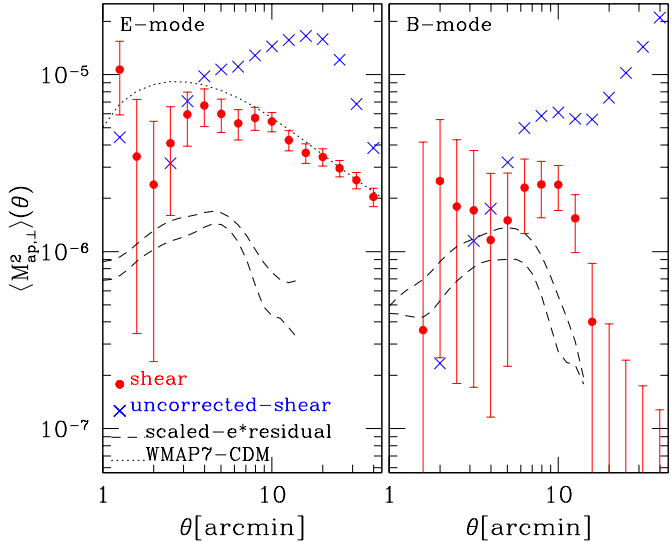}
\end{center}
\caption{Filled circles show E- (left) and B-mode (right) shear aperture
mass variances measured from faint galaxies. 
The error bars present the statistical error computed from 100
randomized realizations. 
The crosses are for {\it uncorrected-shears} (the galaxy shears
to which the correction for the PSF anisotropy is not applied). 
The dashed curves shows variances of the residual PSF ellipticities (but
scaled to the shear by multiplying a factor $(\langle
P_{sm}/P_{\gamma}\rangle /\langle P_{sm}^\ast\rangle)^2\simeq 1.5^2$)
measured from the G5 group of the dense star field data
(see section \ref{sec:PSF} and figure \ref{fig:mapvar_abbgroup_eb.g5cor}); the
upper and lower curves are for $13\times63$ and
$52\times60$-second exposure, respectively.
The dotted curve shows the $\Lambda$CDM prediction based on the 7-year WMAP
cosmological model with the source galaxy redshift distribution inferred
from COSMOS photometric redshift catalog.
\label{fig:mapvar_scana_g5cor}}
\end{figure}

We compute the E- and B-mode aperture mass variance of galaxy shears
through the two-point correlation functions using 
equations (\ref{eq:MapE})-(\ref{eq:xi+-}) with the summations
in equation(\ref{eq:xi+-}) being replaced with a
weighted summation,  $\sum w_i w_j \gamma_i \gamma_j / \sum w_i w_j$,
where $w_i$ is the weight for $i$-th galaxy.
In addition to the shear, we compute the {\it uncorrected-shear}, that
is, the galaxy shear to which the correction for the PSF anisotropy is not
applied.
The results are presented in figure \ref{fig:mapvar_scana_g5cor}.
The error bars indicate the root-mean-square among 100 randomized
realizations, in which the orientations of galaxies are randomized, and
presumably represent the statistical error coming from the galaxy shape
noise. 
It is found from figure \ref{fig:mapvar_scana_g5cor} that the B-mode shear
variance is non-zero on the aperture scale of $\theta\lesssim 10$
arcmin. It should be noted that adjacent bins are correlated.
In order to compare the shear variances with those of the residual PSF
ellipticites, we transform the star ellipticity to the galaxy shear by
multiplying a factor $\langle P_{sm}/P_{\gamma}\rangle /\langle
P_{sm}^\ast\rangle$, which is found to be $\simeq 1.5$ (thus
$\gamma^\ast\simeq 1.5 e^\ast$).
The results from the G5 group of the dense star field data with
the $2\times 2$ sub-region basis correction
(see section \ref{sec:PSF} and figure \ref{fig:mapvar_abbgroup_eb.g5cor}) are
also shown in figure \ref{fig:mapvar_scana_g5cor}.
The upper and lower dashed-curves are for $13\times60$ seconds and $52\times60$
seconds exposure, respectively.
Since the total exposure times of four cosmic shear fields are
$2400-6000$ seconds,  the lower curve could be considered as a ``{\it
  best performance}'' of our PSF ellipticity correction scheme.
The B-mode shear signals at smaller scales  ($\theta\lesssim10$ arcmin)
are larger than this expectation.
A possible reason for this excess is that the stacking of dithered multiple
exposures which is not involved in the dense star field analysis.
This issue will be investigated in detail in a future study in which we
will adopt a PSF correction scheme on an individual exposure basis.
On the other hand, the B-mode variance drops at larger scales
($\theta\gtrsim 15$ arcmin), which is in nice agreement with that
expected from the dense star field analysis.
It is however noticed that in the case of the dense star field data, the
turnaround aperture scale is about 5 arcmin, which is naturally set by
the mean separation of stars ($\sim 1$ arcmin) (see section
\ref{sec:PSFcorrection}), whereas it is about 10 arcmin for the cosmic
shear data.
The reason for this is unclear, though it could be related to an excess
B-mode uncorrected shear variance observed at $\sim10$ arcmin.

%
%
\begin{figure}
\begin{center}
 \includegraphics[height=80mm]{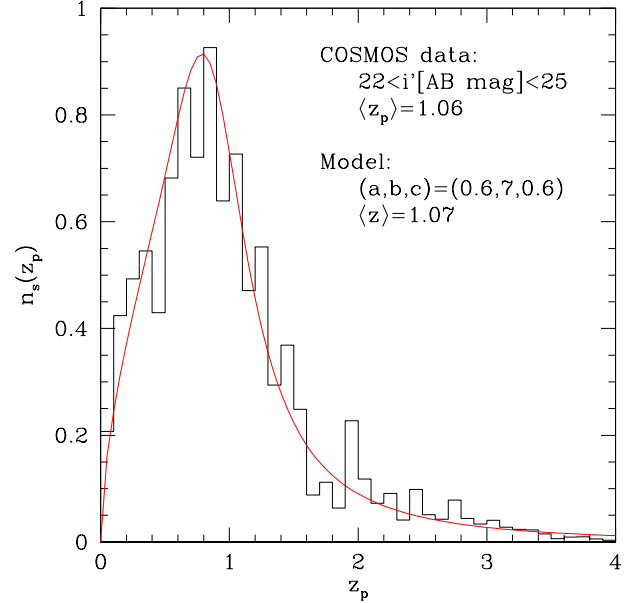}
\end{center}
\caption{Redshift distribution of galaxies used for the cosmic shear
  analysis derived adopting the COSMOS photometric redshift catalog
  (Ilbert et al 2009).
 The solid curve shows the fitting model by equation \ref{eq:nz} with
 the parameters denoted in the panel.
The mean redshift of the model is 1.07 which is in a very good agreement
with the value derived from the photometric redshift data ($\langle
z_p\rangle =1.06$).
\label{fig:nsw}}
\end{figure}

Finally, we evaluate the influence of the residual PSF anisotropies on
the cosmic shear E-mode measurement.
We plot in the left panel of figure \ref{fig:mapvar_scana_g5cor} the
$\Lambda$CDM prediction based on the 7-year WMAP cosmological model
(Komatsu et al. 2011) as a basis for comparison. 
In computing the $\Lambda$CDM prediction, we used the {\sc halofit} model by
Smith et al. (2003), and we used the redshift
distribution of source galaxies inferred by the following manner:
We utilized the COSMOS field data. 
We merged the galaxy catalog used for the cosmic shear
measurement with the COSMOS photometric redshift catalog (Ilbert et al
2009), and compute the redshift distribution by adopting the photometric
redshift, which is presented in  figure \ref{fig:nsw}.
We fit the redshift distribution with the following function (Fu et
al. 2008
\begin{equation}
\label{eq:nz}
n(z)=A {{z^a+z^{ab}}\over {z^b + c}},
\end{equation}
where the normalization, $A$, is determined by imposing $\int dz~n(z) =1$
within the integration range $0<z<6$.
We found that a parameter set $(a,b,c)=(0.7,~6,~0.7)$ gives a reasonably
good fit to the data, as shown in figure \ref{fig:nsw}.
It is found from a comparison between the $\Lambda$CDM prediction and
the residual PSF shear variance evaluated from the dens star field
analysis (figure  \ref{fig:mapvar_scana_g5cor}) that on scales below 10
arcmin, the residual shear variance is more than 10 percents of
the expected cosmic shear variance.  
The comparison between the measured E- and B-mode shear variances on
those scales results in even worse figures.
On the other hand, on larger scales ($\theta>15$ arcmin), where the
anisotropic PSF correction works well, the residual variance is well
less than 10 percents of the measured and expected E-mode variance.
This is one of important findings of this paper.
Before closing, we comment on an effect of the noise bias 
(Refregier et al. 2012; Melchior \& Viola 2012; Okura \& Futamase 2012)
on the above analysis. 
The noise bias introduces a systematically lower ellipticity value
with its amplitude depending on the SN (a smaller bias for a higher SN
object).
Refregier et al. (2012) found that the amplitude of noise bias is a few
percent for SN$=10$ objects. 
Since the threshold SNs adopted in the above analysis are SN$>10$ for the
cosmic shear galaxy selection and SN$>40$ for the dens star field analysis,
the expected amplitude of the noise bias on the above shear
variance measurements might be less than 5 percent. 
Therefore, conclusions from the above analysis are, at least
qualitatively, not affected by the noise bias.

\section{Summary and discussion}
\label{sec:summary}

We analyzed the anisotropic PSF of Suprime-Cam data utilizing
the dense star field data. 
We decomposed the PSF ellipticities into three components (the optical
aberration, atmospheric turbulence, and the chip-misalignment) in an
empirical manner, and assessed the amplitude of each component.
We then tested a standard method for correcting the PSF ellipticities
against a mock simulation based on models of the PSF ellipticities
obtained from the dense star field data analysis. 
Finally we examined the impact of the residual PSF ellipticities on
the cosmic shear measurement.

We found that the optical aberration has the largest contribution to the
PSF ellipticities of long-exposure (say, longer than 10 minutes) data.
We also found that the spatial variation of the optical PSFs can be
modeled by a simple three-component model, which is based on the
lowest-order aberration theory described in appendix
\ref{appendix:aberration}. 
It was also found that the optical PSF ellipticities vary smoothly
on the focal plane, and thus are well corrected by the standard correction
method in which the spatial variation is modeled by a polynomial
function. 

The PSF ellipticities after subtracting the optical PSF model ellipticities
show discontinuities between CCD chips, which indicates that the
misalignment of chips on the focal plane induces an additional PSF
ellipticity. 
In order to make a crude estimation of the amplitude of the 
chip-misalignment PSF ellipticities, we fitted the residual PSF
ellipticities to the 2nd order bi-polynomial function on a chip-by-chip
basis.
It turns out that the amplitude of the chip-misalignment PSF
ellipticities is lower than the optical component, though it is not
negligible. 
It was found from the mock simulation that the chip-misalignment
component combined with the optical component puts a lower limit on the
capability of the PSF correction, which can be, in principle, avoided by
employing a chip-basis correction scheme.

We investigated the properties of PSF ellipticities resulting from the
atmospheric turbulence using a numerical simulation of wave
propagation through atmospheric turbulence under a typical weather
condition of the Subaru telescope at Mauna-kea (appendix 
\ref{appendix:atmos}).
{}From the simulation results, we evaluated the power spectrum and
aperture mass variance of the atmospheric PSF ellipticities, and 
derived power-law fitting functions of them.
As was already pointed out in literature (Wittman 2005; de Vries et al
2007; Jee \& Tyson 20011; Chang et al. 2012; 2013 Heymans et al. 2012),
the RMS amplitude of the atmospheric PSF ellipticities decreases as
$T_{exp}^{-1/2}$.
We computed the aperture mass variance of the dense star field data for
various exposure time (from 30-second to $52\times60$-second).
The amplitude of the atmospheric PSF ellipticities was evaluated by
assuming that the deepest data represents the amplitude of
(quasi-)static PSF components, and the difference from that is due to
the atmospheric PSFs ellipticities.
The results are found to be in reasonable agreement with the
simulation results.
Therefore, we may conclude that the aperture mass variance of the
atmospheric PSF ellipticies for a long-exposure data (say, longer than
10 minutes), is at least one order of magnitude smaller than that of the
optical PSF ellipticities. Since the atmospheric PSFs are not smoothly
varying component, its correction is affected by the common limitation
that the PSF correction on scales smaller than the mean star separation
works very poorly because of the poor sampling of the spatial variation of
PSFs on those scales.

The above findings provide us a clue to develop an optimal PSF interpolation
scheme,
It is found that the spatial variation of PSF ellipticities consists of
two components: one is a smooth and parameterizable component
arising from the optical aberration and chip-misalignment; the other
is a non-smooth and stochastic component arising from the atmospheric
PSFs. 
The former can be modeled with a parametric model, as shown in this
paper. Also it has been argued that an interpolation scheme based on the
principle-component analysis is effective for such a case (Jarvis \&
Jain 2004; Jee \& Tyson 2011; see also Miyatake et al 2013 and Lupton et
al 2001 for an actual implementation for Suprime-Cam data reduction
pipeline).
On the other hand, it is shown in Berg\'e, et al (2012) and
Gentile et al. (2013) that local-type correction schemes, such like the
radial basis functions and Kriging work well for atmospheric PSFs.
Apparently, a hybrid interpolation scheme, in which the above two types
of interpolation schemes are optimally incorporated, is a strong
candidate for achieving a better PSF correction.

We examined the effects of the residual PSF anisotropies on
Suprime-Cam cosmic shear data.
We also compared the B-mode shear variance measured from 5.6-degree$^2$
$i'$-band data with the residual PSF ellipticities (but being properly
transformed into shear) measured from the dense star field data, which
can be considered as the ``{best performance}'' of our PSF ellipticity
correction scheme. 
It is found that the shape and amplitude of the B-mode shear variance are 
broadly consistent with those of the residual PSF ellipticities.
This indicates that most of the sources of residual systematic are
understood, which is an important step for cosmic shear
statistics to be a practical tool of precision cosmology.
However, it is also found that the B-mode shear amplitude at scales $\sim
10$ arcmin are systematically larger than the residual PSF.
The reason for this excess is unclear; one possible reason is thestacking
of dithered multiple exposures, which is not involved in the dense star
field analysis.
Such stacking-related issues may be avoided by employing a weak-lensing
shape measurement scheme on the basis of individual exposures (e.g.,
Miller et al. 2007; 2013; and Miyatake et al. 2013), which combined with
a hybrid interpolation scheme will be addressed in a future work.

\bigskip

We thank M. Takada, M. Oguri and H. Miyatake for useful discussions.
We thank Y. Utsumi for help with the photometric calibration of
Suprime-Cam data.
We also thank L. Van Waerbeke, H. Hoekstra J. D. Rhodes and
R. Mandelbaum for valuable comments on an earlier manuscript, which
improved the paper. 
TH greatly thanks Y. Mellier for many fruitful discussions on various
aspects of this work.
We would like to thank M. Britton for making the {\sc Arroyo} software
available.
Numerical computations in this paper were in part carried out on the
general-purpose PC farm at Center for Computational Astrophysics, CfCA,
of National Astronomical Observatory of Japan.
This work is based in part on data collected at Subaru Telescope and
obtained from the SMOKA, which is operated by the Astronomy Data Center,
National Astronomical Observatory of Japan.
This work is supported in part by Grant-in-Aid for
Scientific Research from the JSPS Promotion of Science
(23540324).

\appendix

\section{PSF ellipticity originated from the third-order optical aberrations}
\label{appendix:aberration}

\begin{figure}
\begin{center}
\includegraphics[width=40mm]{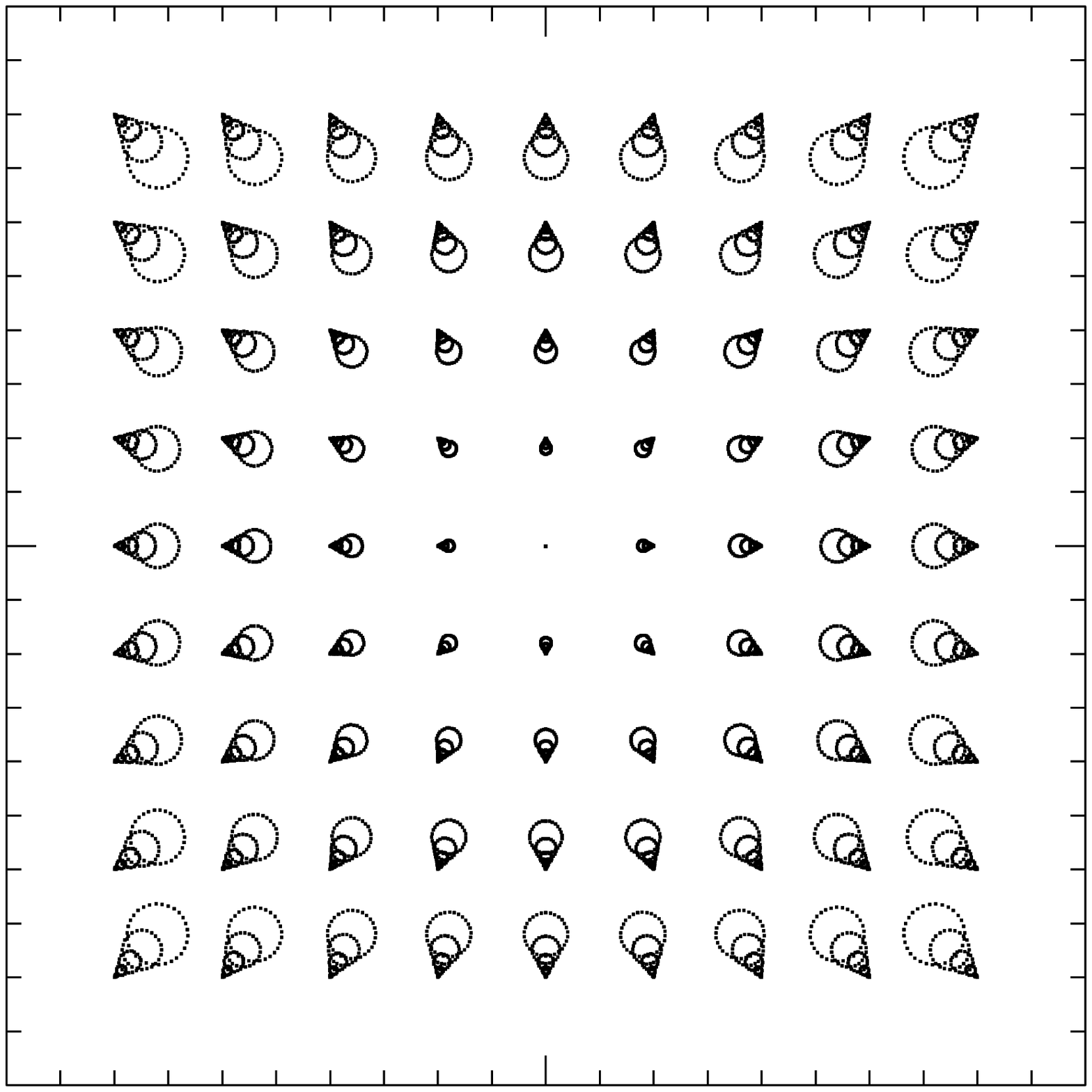}
\includegraphics[width=40mm]{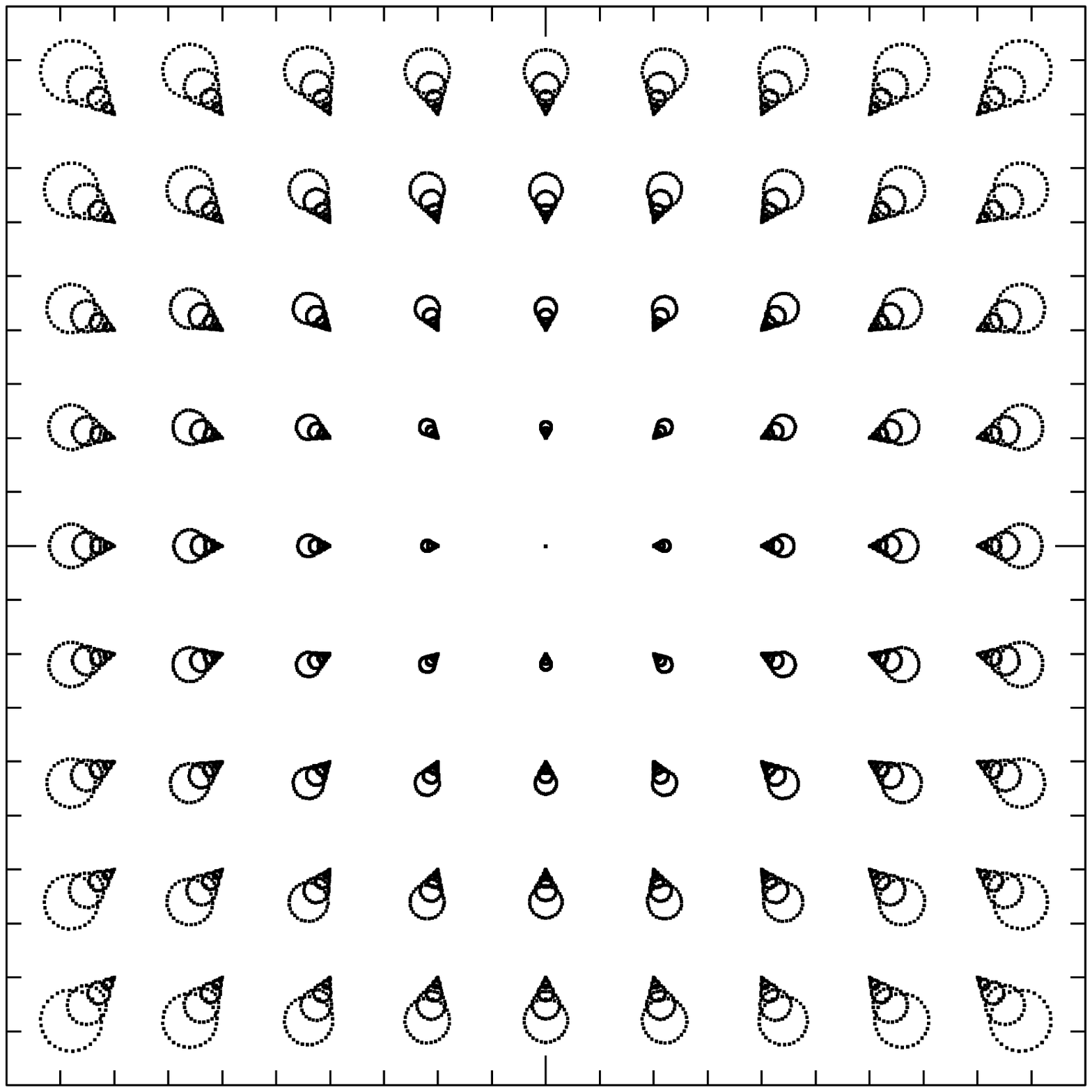}\\
\vspace{1mm}
\includegraphics[width=40mm]{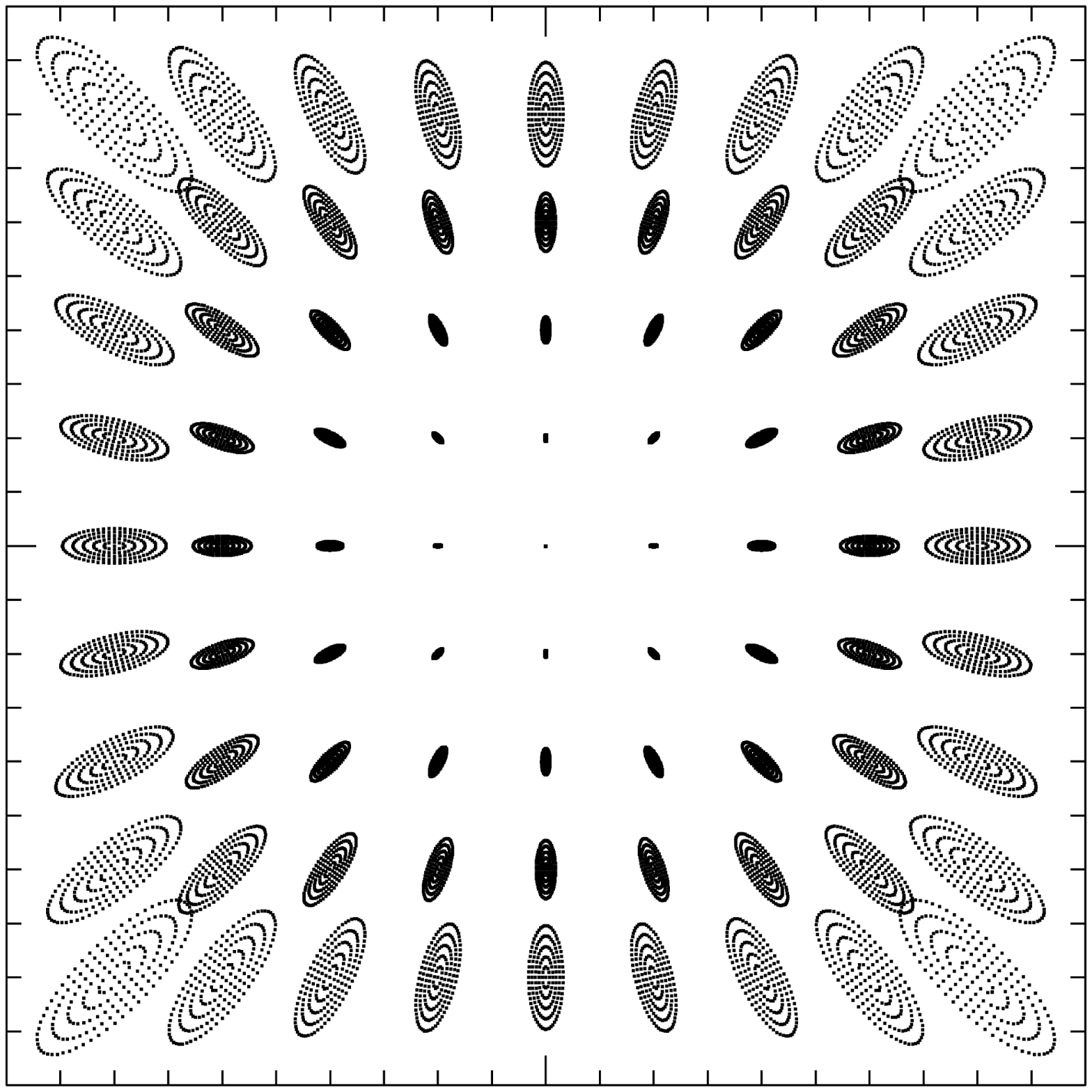}
\includegraphics[width=40mm]{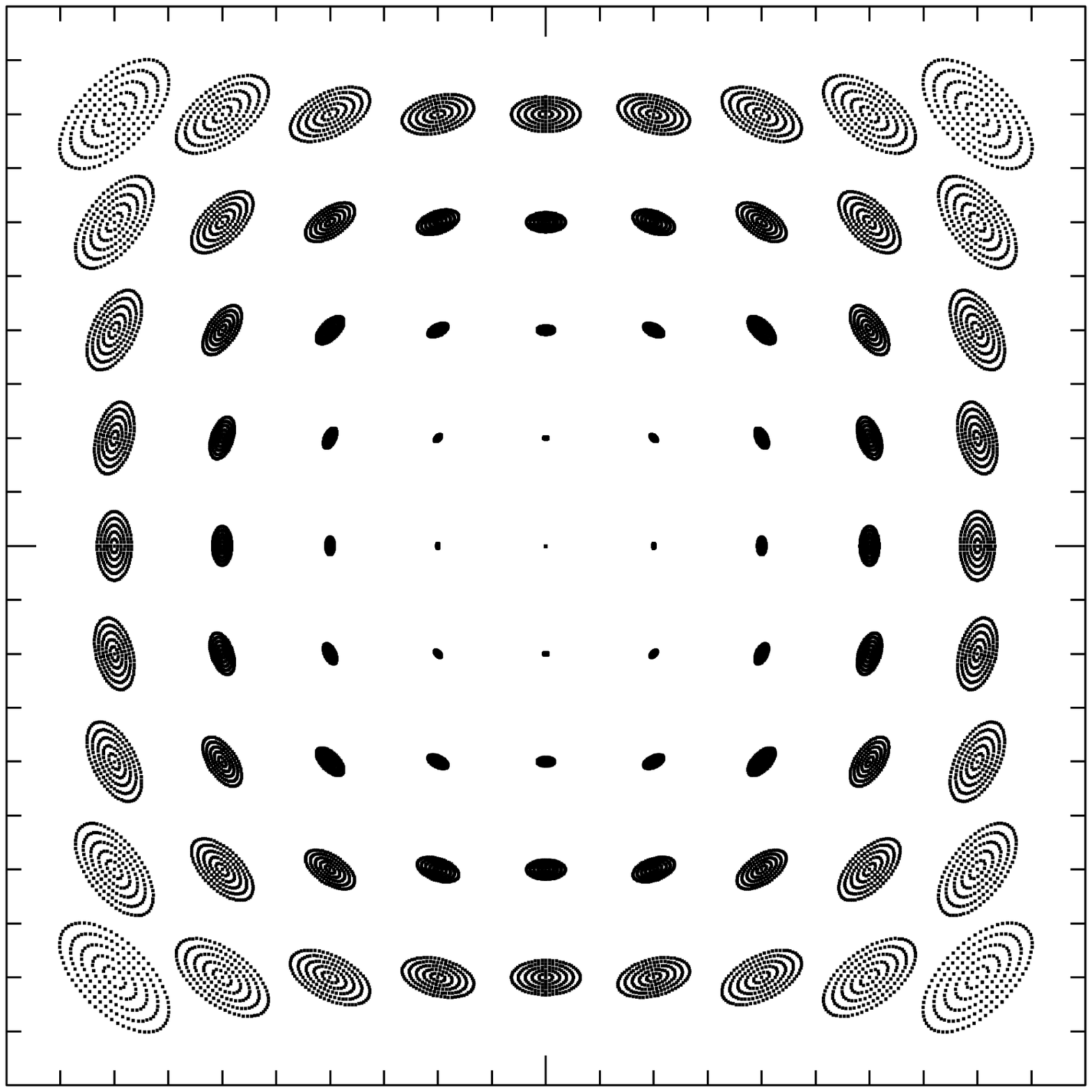}\\
\vspace{1mm}
\end{center}
\caption{{\it Top two panels}: spot diagrams for the Seidel coma with
  opposite signs of $W_{131}$. In both cases, the major axis of the
  ellipticity of the PSF are radially oriented with a scaling of
  $|e|\propto r^2$. 
  {\it Bottom two panels}: spot diagrams for a combination of
  astigmatism and field curvature: Bottom-left panel is for a case for
  $|W_{222}+W_{220}|>|W_{220}|$, whereas the bottom-right panel is for  
$|W_{222}+W_{220}|<|W_{220}|$. In both cases, the ellipticity parameter
  is scaled as $|e|\propto r^4$.
\label{fig:coma}}
\end{figure}

We summarize the properties of ellipticity in the PSF caused by 
third-order optical aberrations with and without rotational symmetry.
We consider a system with a circular pupil.
We follow the notation by Thompson (2005, see Fig 1 of his paper):
$\vec{H}$ represents the position in the image field; without loss of
generality we can choose $(H_x,H_y)=(0,h)$. $\vec{\rho}$ represents the
position in the exit pupil with $(\rho_x,\rho_y)=(\rho \sin\phi, \rho
\cos\phi)$. 

Let us start with a rotationally symmetric system. In this case, the
wavefront distribution is given by (Thompson 2005)
\begin{eqnarray}
\label{eq:seidel}
W&=&W_{040}\rho^4 + W_{131}h\rho^3\cos\phi + W_{222}h^2\rho^2\cos^2\phi \nonumber\\
&&+ W_{220}h^2\rho^2 + W_{311}h^3\rho\cos\phi.
\end{eqnarray}
Among those five Seidel aberrations, those that make an ellipticity parameter
$e$ non-zero are the Seidel coma (the second term) and astigmatism (the
third term).
The Seidel coma results in an ellipticity whose major axis oriented
toward radial direction with $|e| \propto r^2$ where $r$ is the 
distance from the center of the image field.
On the other hand, a combination of the astigmatism and the field curvature
(the fourth term) results in a radially elongated (if
$|W_{222}+W_{220}|>|W_{220}|$) or tangentially elongated (if
$|W_{222}+W_{220}|<|W_{220}|$) ellipticity
parameter with $|e|\propto r^4$ (see figure \ref{fig:coma}).
The last term, the distortion, does not generate ellipticity for a
point source because it only introduces a shifting of the image position on a
focal plane. However, for an extended source, its differential effect
causes a radially or tangentially elongated ellipticity with $|e|\propto
r^2$. It should, however, be noticed that unlike other aberration terms,
the image deformation caused by the distortion is due to mapping, and thus
this should be treated separately from the PSF.

\begin{figure}
\begin{center}
\includegraphics[width=40mm]{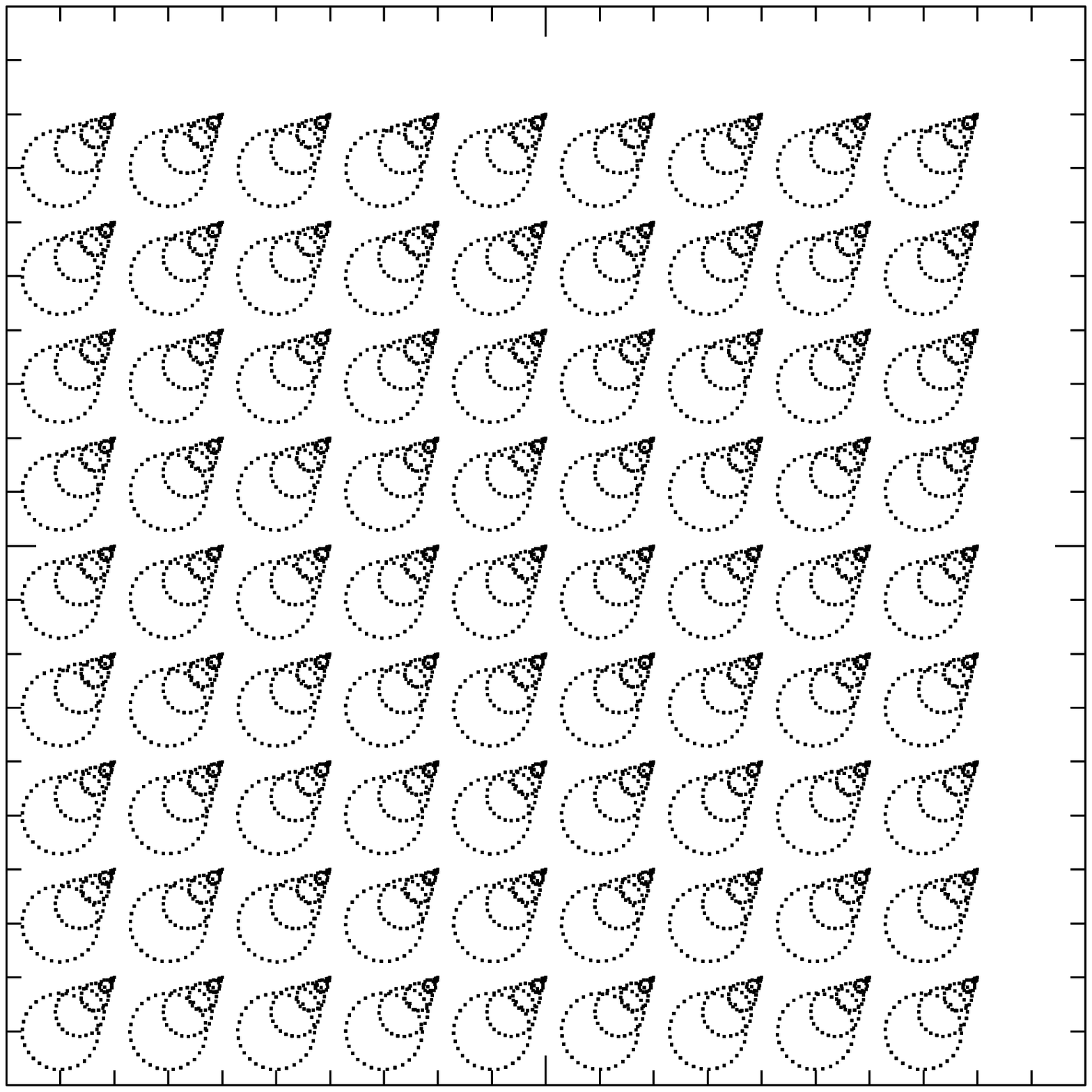}
\includegraphics[width=40mm]{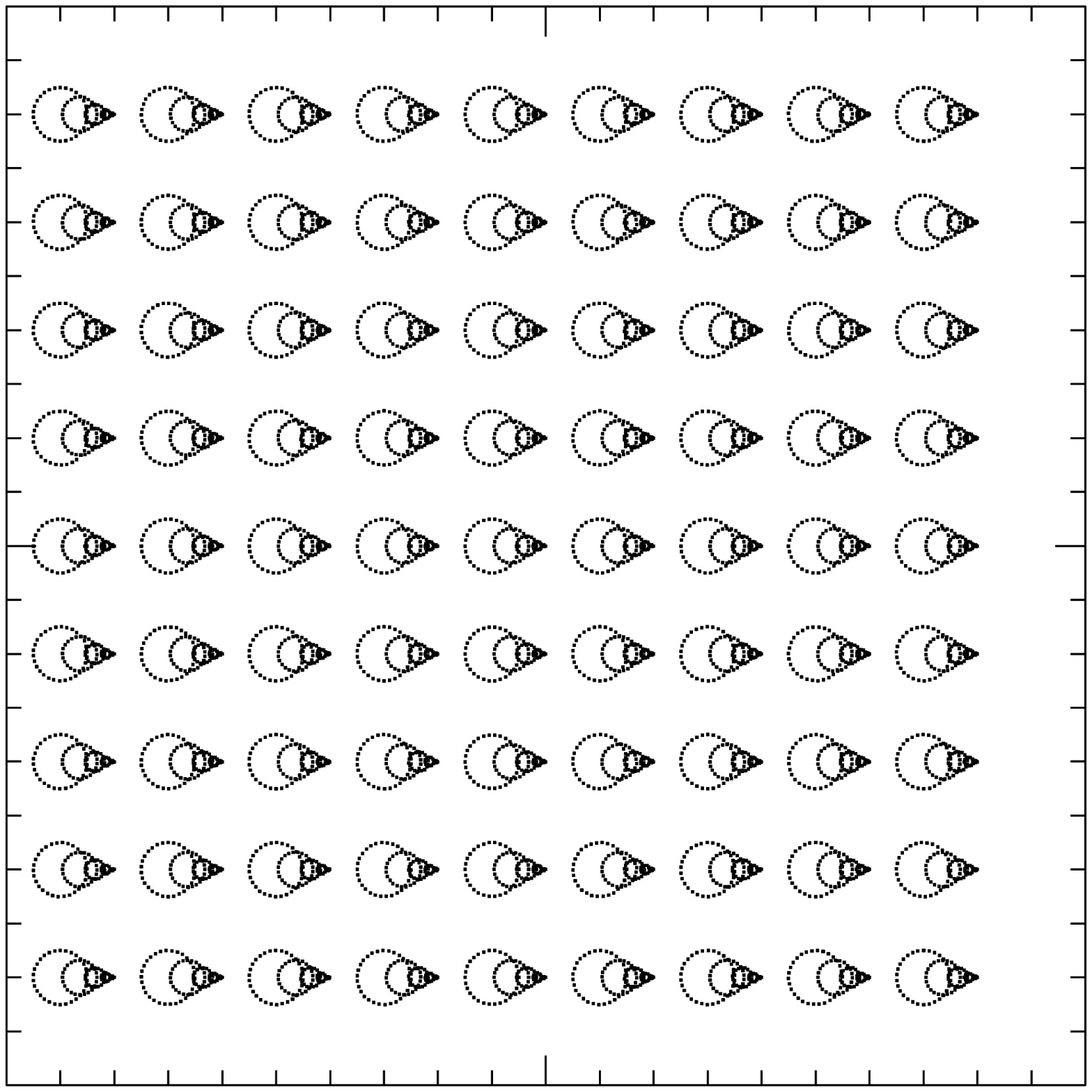}\\
\vspace{1mm}
\includegraphics[width=40mm]{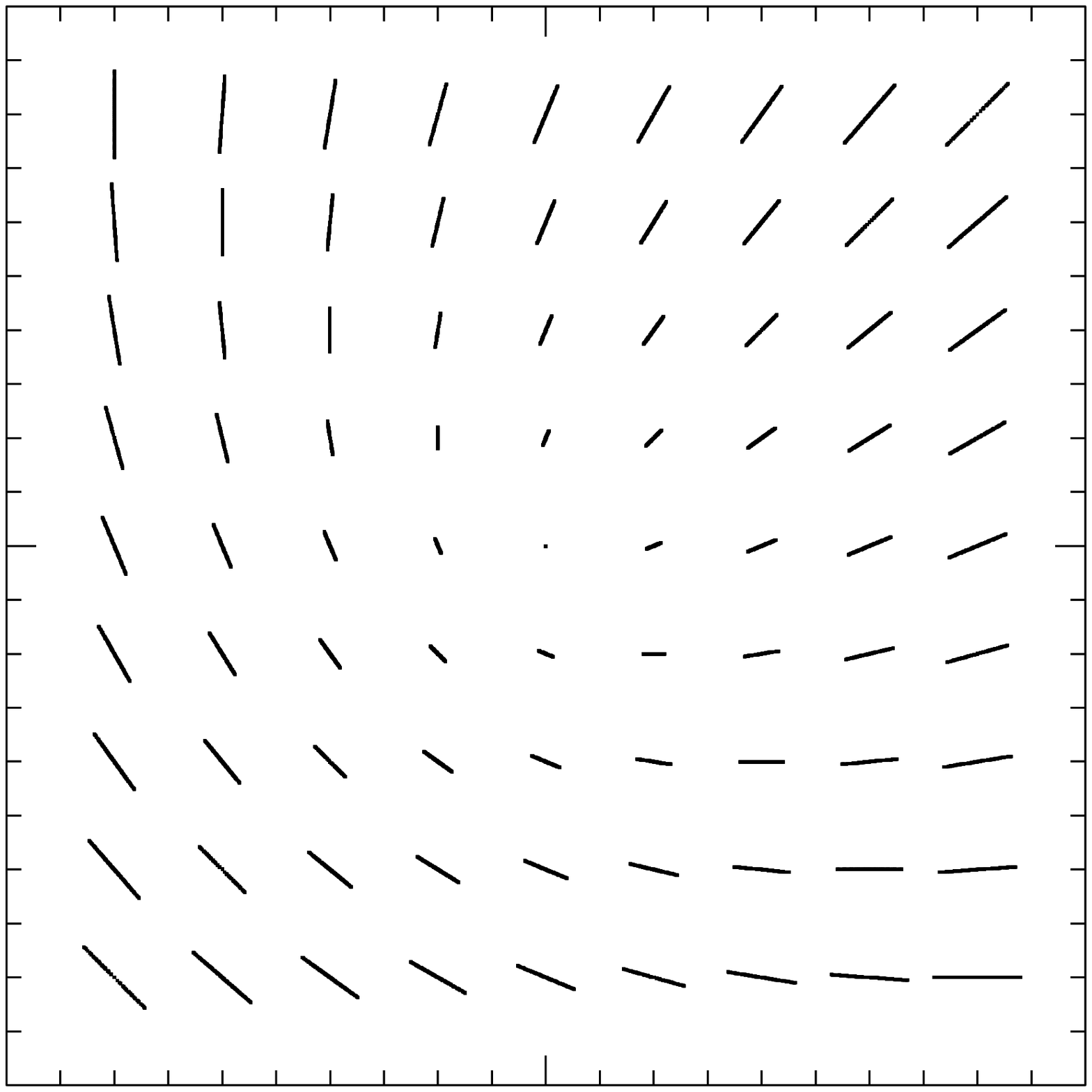}
\includegraphics[width=40mm]{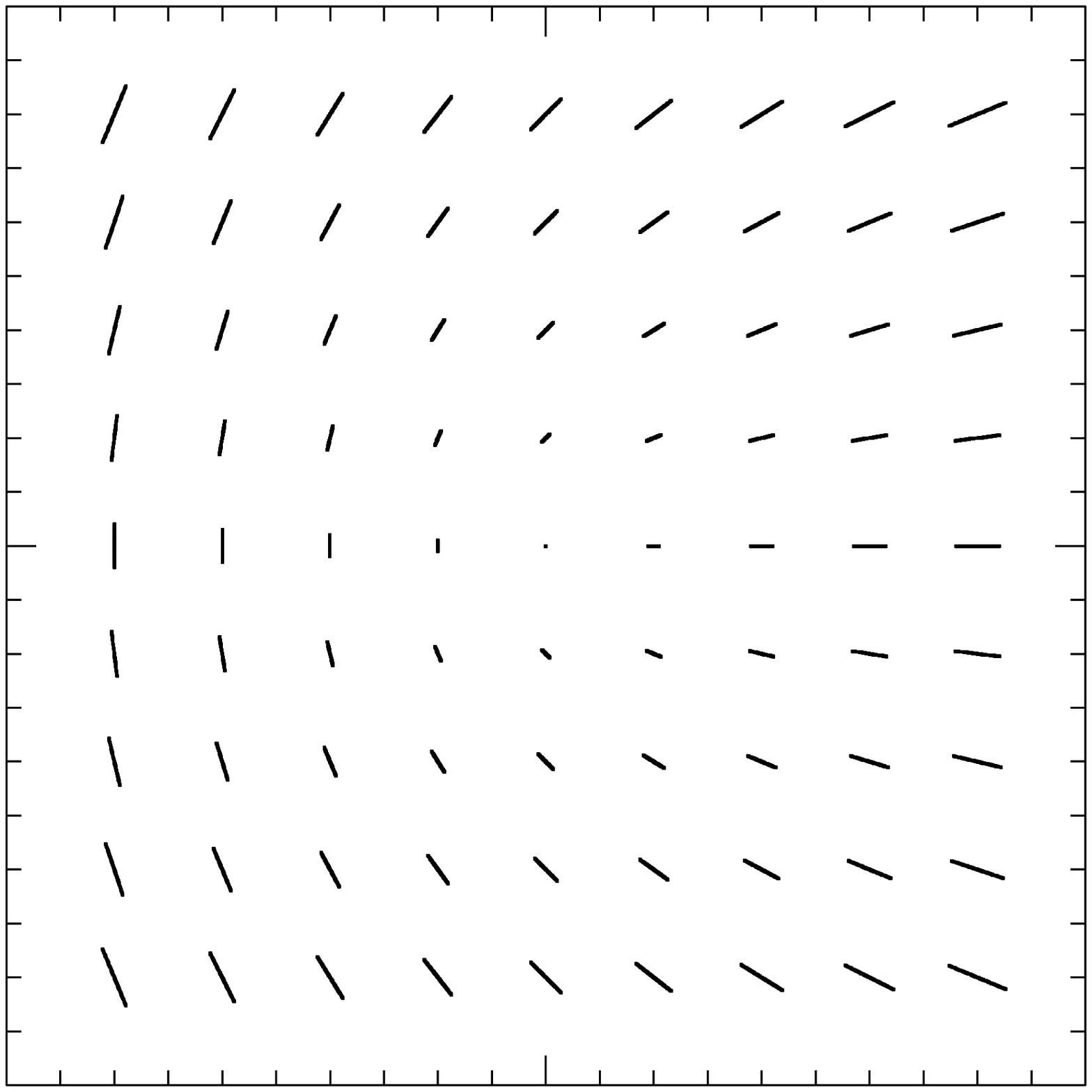}\\
\vspace{1mm}
\end{center}
\caption{{\it Top two panels}: spot diagrams for the misalignment coma with
  different misalignment directions.
  {\it Bottom two panels}: spot diagrams for the misalignment
  astigmatism with different misalignment directions.
In both cases, the ellipticity parameter
  is scaled as $|e|\propto r^2$.
\label{fig:mis}}
\end{figure}

Next, we consider the case without rotational symmetry, allowing for
a tilt and decenter between the axes of the exit pupil and the image field.
To compute the wavefront distribution using a perturbative approach, it is
customary to introduce a vector $\vec{\sigma}$ with
$(\sigma_x,\sigma_y)=(\sigma\sin \theta,\sigma\cos \theta)$ which
represents the decentration of the center of the aberration field, $W$,
with respect to the unperturbed field center. 
In this case, optical aberrations result in a PSF with non-zero
ellipticity parameter are the {\it misalignment} coma and the {\it
  misalignment} astigmatism 
(Schechter \& Levinson 2011).

The wavefront distribution for the misalignment coma
is given by
\begin{eqnarray}
\label{eq:mis-coma}
W \propto \sigma \rho^3\cos(\phi-\theta).
\end{eqnarray}
In this case, the PSF has the same coma shape as the Seidel coma.
However, the PSF does not depends on the position in the image field, since
the above expression does not include $h$, but the PSF (and thus the
ellipticity) is constant over the image field, as shown in the two top panels
of figure \ref{fig:mis}.

The wavefront distribution for the misalignment astigmatism is given by
\begin{eqnarray}
\label{eq:mis-mis}
W \propto \sigma h \rho^2 \cos^2(\phi-{\theta / 2}).
\end{eqnarray}
In this case, the ellipticity of the PSFs depends not only on the distance
from the center of the image field, but also on the azimuth
angle (we shall denote it by $\psi$), and is written by
$(e_1,e_2)\propto (r^2\cos(\psi-\theta),r^2\sin(\psi-\theta))$ 
(see figure \ref{fig:mis}).

Although an actual optical aberration may result in a more complicated
PSF than that expected from the third-order optical aberrations, it may
be reasonable to build a model of the PSF ellipticity caused by an optical
aberration based on the above results.
We consider the following form of a model (already introduced in 
equation (\ref{eq:PSFaberrat}), but rewritten for self-contained) that
consists three terms, namely the constant, axis-symmetric and asymmetric
components; 
\begin{eqnarray*}
\left(
\begin{array}{c}
  e_1\\ e_2
\end{array}
\right)
&=&
\left(
\begin{array}{c}
  c_1\\ c_2
\end{array}
\right)
+
\sum_{n_s}s_{n_s} r^{n_s} 
\left(
\begin{array}{c}
  \cos2\psi \\ \sin2\psi
\end{array}
\right)\nonumber \\
&&+
\sum_{n_a}a_{n_a} r^{n_a} 
\left(
\begin{array}{c}
  \cos(\psi-\theta)\\ 
  \sin(\psi-\theta) 
\end{array}
\right).
\end{eqnarray*}
We determine the model parameters $s_{n_s}$, $a_{n_a}$ and $\theta$ by 
the standard least square method with the observed PSF ellipticities.
In doing so, we take $n_s=\{1,2,3,4\}$ and $n_a=\{1,2\}$.
As shown in section \ref{sec:PSF}, this simple model succeeds very well
in fitting the global pattern of PSF ellipticities in Suprime-Cam data.

\section{A numerical study of PSF ellipticities caused by atmospheric
  turbulence}
\label{appendix:atmos}

We examined the statistical properties of PSF ellipticities caused by
atmospheric truculence using a numerical simulation. 
The aim of this study is to investigate the shape and amplitude of the
power spectrum of atmospheric PSF ellipticities, which is essential 
knowledge for understanding the statistical properties of PSF ellipticities.
Our approach is similar to one taken by Jee \& Tyson (2011), that is to
utilize a numerical simulation of wave propagation through atmospheric turbulence
for evaluating atmospheric PSFs.

Here, we briefly describe our numerical simulation.
Since we heavily use publicity open software, {\sc Arroyo} (Britton
2004), we refer the reader to the above reference and software
documents\footnote{http://eraserhead.caltech.edu/arroyo/arroyo.html}
for details of models, computational methods and their implementation. 
The atmospheric turbulence was modeled by eleven-layer frozen
Kolmogorov screens with the atmospheric model at Mauna-Kea
proposed by Ellerbroek \& Rigaut (2000).
The strength of the total atmospheric turbulence was set by the Fried
length, $r_0$, for which we adopt a value proposed by Ellerbroek \& Rigaut
(2000): $r_0=0.23$m at a wavelength of $0.5\mu$m. 
Neither a large scale nor small scale cutoff (the, so-called, outer and
inner scale) on the Kolmogorov power spectrum is imposed.
The wind model given by equation (3.20) of Hardy (1998) was adopted with
random wind directions.
Having the atmospheric model ready, an atmospherically disturbed phase
function of a wavefront is computed.
We consider a monochromatic electromagnetic wave with a wavelength of
$0.8\mu$m. 
Then, an instantaneous PSF was obtained by Fourier transforming the phase
function within a pupil, for which we assumed a circular pupil
with 8.2m diameter for simplicity. 
All of the above models and computations were implemented in {\sc Arroyo}
subroutines.
Finally, a sequence of instantaneous PSFs were added to obtain a
long-exposure PSF.
We computed PSFs of 1, 10 and 60 seconds exposures on $64\times64$
regular grids over a $30\arcmin \times30\arcmin$ focal plane. 

\begin{figure}
\begin{center}
\includegraphics[height=88mm,angle=-90]{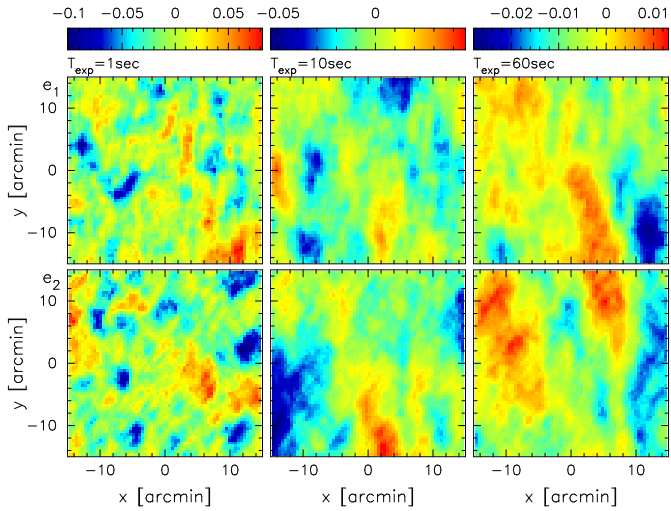}
\end{center}
\caption{Ellipticity magnitude obtained from the numerical simulation
  described in appendix \ref{appendix:atmos}: upper and
  lower panels are for $e_1$ and $e_2$, respectively, and the color
  scales are displayed on the top of those panels.
The exposure times are, from left to right, 1-second, 10-second and 60-second,
respectively.
\label{fig:emap2x3.run105}}
\end{figure}

We generated 20 realizations. The mean FWHM of PSFs among the
realizations was 0.65 arcsec for a 60-seconds exposure, which is in a good
agreement with the observed value at the Subaru telescope (Miyazaki et
al. 2002). 
AS an illustrative example, one realization is shown in Figure
\ref{fig:emap2x3.run105}, where the ripple like feature appears.
Similar features were observed in CFHT data (Heymans et al. 2012) and
Suprime-Cam data (see figures \ref{fig:emap1} and \ref{fig:emap2}).
It should, however, be noted that the appearance varies widely between
realizations, depending on turbulence and wind properties.

\begin{figure}
\begin{center}
\includegraphics[width=80mm]{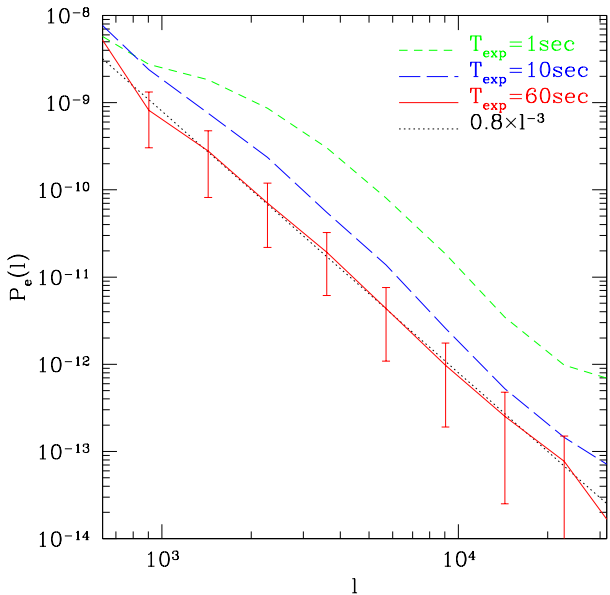}
\end{center}
\caption{Power spectrum of PSF ellipticities measured from
  the simulation results.
The dashed, long-dashed and solid lines are for 1-second, 10-second and 60-second
exposures, respectively.
The error bars, plotted only for the 60-second result for clarify, show the
RMS among 20 realizations. 
The dotted line shows the power-law model of $P_e(l)=0.5\times l^{-3}$
which turns out to be a good approximation of the 60-second result.
\label{fig:powerspec_arroyo}}
\end{figure}

\begin{figure}
\begin{center}
\includegraphics[width=80mm]{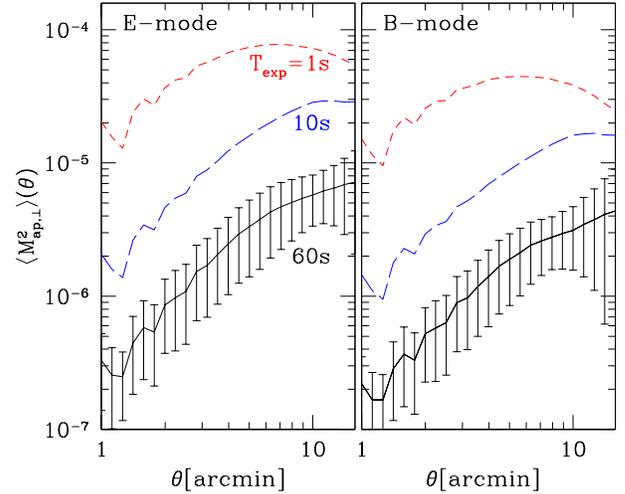}
\end{center}
\caption{Aperture mass variance of PSF ellipticities measured from
  the simulation results.
The dashed, long-dashed and solid lines are for 1-second, 10-second and 60-second
exposures, respectively.
The error bars, plotted only for the 60-second result for clarify, show the
RMS among 20 realizations. 
\label{fig:mapvar_arroyo}}
\end{figure}

We measured the power spectrum and the aperture mass variance from the
simulation data and present them in figures \ref{fig:powerspec_arroyo} and
\ref{fig:mapvar_arroyo}, respectively.
The following three major findings were derived from the figures:
First, the amplitude of the power spectrum decreases as $P_e\propto
T_{exp}^{-1}$, as was already pointed out in previous studies (Wittman
2005; de Vries et al 2007; Jee \& Tyson 20011; Chang et al. 2012; 2013
Heymans et al. 2012).  
Second, the power spectrum is approximated by a power-law model with a power
index of $-3$.
A plausible reason for the power-law shape is our assumption of
the power-law turbulence power spectrum without a cutoff.
In reality, a large-scale cutoff (the, so-called, outer scale) at around
10$-$100m is expected (e.g., Hardy 1998), which may result in a cutoff in
the PSF ellipticity power spectrum at a scale of around 1 degree.
Third, the atmospheric turbulence results in E/B-mode almost equally
partitioned PSF ellipticities. 
Through a closer look at figure \ref{fig:mapvar_arroyo}, one may however find
that the E-mode amplitude is slightly, but systematically, larger than
the B-mode. The reason for this is unclear and we leave it for future work.

In addition to the above findings, we also find from
figures \ref{fig:powerspec_arroyo} and \ref{fig:mapvar_arroyo} that
models $P_e(l)\sim 0.8l^{-3}$ and  
$\langle M_{ap,\perp}^2 \rangle (\theta) \sim 4\times10^{-7} \theta$
give a good fit to the 60-second result.
In what follows, we discuss how those correlation amplitudes depend on
other observational parameters, including the exposure time and the
atmospheric seeing FWHM ($\theta_{\rm atm}$).
Consider the case $D\gg r_0$, where $D$ denotes the diameter of a telescope,
and a monochromatic radiation with a wavelength $\lambda$ from a point
source. The wavefront from a point source is disturbed by atmospheric turbulence,
and we consider, for a simple approximation, the disturbed wavefront to be
segmented $r_0$-sized patches of constant phase and random phases
between the individual patches (Saha 2007, and see e.g., figure 5.3 of that
textbook for an illustration).
The number of patches within a pupil is approximately $N_{patch}\sim (D/r_0)^2$.
A wavefront from each patch results in a sharp speckle PSF on a focal plane,
and a total PSF is viewed as a superposition of randomly distributed
speckles over an extent of the atmospheric seeing size 
$\theta_{\rm atm}\sim \lambda/r_0$ (Hardy 1998; Saha 2000; see for 
illustrative examples, figure 3 of Kaiser, Tonry \& Luppino 2000; and Fig 5
of Jee \& Tyson 2011). 
In this picture, the PSF anisotropy is understood as a natural
consequence of a finite number of speckles.
Since the disturbed wavefront changes quickly, more different speckles
are accumulated as $N_{spec}\propto T_{exp}$, and thus the PSF becomes
rounder. 
It is found from the above numerical simulation that the RMS of PSF
ellipticities decreases as $\langle e^2 \rangle^{1/2}\propto T_{exp}^{-1/2}$, which
combined with the above mentioned relationships leads to the following
relationships; 
\begin{eqnarray}
\langle e^2 \rangle^{1/2}
&\propto& N_{spec}^{-1/2} 
\propto T_{exp}^{-1/2} N_{patch}^{-1/2} \nonumber \\
&\propto& T_{exp}^{-1/2} r_0 \propto
T_{exp}^{-1/2}/\theta_{\rm atm}.\nonumber
\end{eqnarray}
We tested this relationship against numerical simulations with different
values of $r_0$, from which the above dependences of $r_0$ and 
$\theta_{\rm atm}$ on $\langle e^2 \rangle^{1/2}$ were verified.
Combined the last relationship with the power-law fitting models, 
we have
\begin{eqnarray}
  \label{eq:ps-scaling}
  P_e(l)&\sim& 0.8\times l^{-3}
\times \left({{T_{exp}}\over {1{\rm  min}}}\right)^{-1} 
 \left({\theta_{\rm atm}\over {0.65{\rm  arcsec}}}\right)^{-2},
\end{eqnarray}
and
\begin{eqnarray}
  \label{eq:map-scaling}
  \langle M_{ap,\perp}^2 \rangle (\theta)
&\sim& 4\times10^{-7}\times \theta \nonumber \\
&\times&\left({{T_{exp}}\over {1{\rm  min}}}\right)^{-1} 
 \left({\theta_{\rm atm}\over {0.65{\rm  arcsec}}}\right)^{-2}.
\end{eqnarray}
Although the above relationships are quite crude,
they are useful to quickly evaluate a magnitude of atmospheric PSF
ellipticities for a given set of observational parameters.
Note that these relationships are valid only for a telescope
with a diameter of 8.2m.
It may be worth pointing out that it is found from numerical simulations
with different telescope diameters that the amplitude of PSF
ellipticities is insensitive to $D$.
This differs from a scaling relationship of $P_e \propto D^2$ expected
from the above consideration (likewise for $\langle M_{ap,\perp}^2 \rangle $).
This difference may be explained by the diameter dependence of the PSF,
which is not taken into account in the above consideration.
According to the diffraction theory (e.g., Hardy 1998; Saha 2007), the
PSF is the inverse Fourier transform of the optical transfer function
(OTF) which is the auto correlation function of the product of the
telescope pupil function with the atmospheric phase function. 
In that relationship, the diameter of telescope enters through the pupil
function. Therefore, roughly and intuitively speaking, the power
spectrum of PSF ellipticities relates to the OTF, and the diameter acts
as a large-scale cutoff. 
Since the turbulence power spectrum has a larger power on a larger
scales (Kolmogorov power spectrum has the shape of $k^{-11/3}$), the
large-scale cutoff imposed by the telescope diameter may have a great
impact on the amplitude of the PSF ellipticities.


\end{document}